\documentclass[fleqn,usenatbib]{mnras}

\usepackage{newtxtext,newtxmath}

\usepackage[T1]{fontenc}
\usepackage{ae,aecompl}

\usepackage{subfig}
\usepackage{tikz}
\usepackage{standalone}

\usepackage{graphicx}	
\usepackage{amsmath}	
\usepackage{amssymb}	
\usepackage{booktabs}

\def\beq{\begin{eqnarray}}
\def\eeq{\end{eqnarray}}

\newcommand{\av}[1]{\langle{#1\rangle}} 
\let\vec\mathbf

\newcommand{\resub}[1]{{#1}}

\numberwithin{equation}{section}

\title[RascalC: Estimating Galaxy Covariance Matrices]{RascalC: A Jackknife Approach to Estimating Single and Multi-Tracer Galaxy Covariance Matrices}

\author[O.\,H.\,E.\,Philcox et al.]{
Oliver H.\,E. Philcox,$^{1}$\thanks{E-mail: \href{mailto:ohep2@alumni.cam.ac.uk}{ohep2@alumni.cam.ac.uk}}
Daniel J. Eisenstein,$^{1}$
Ross O'Connell$^{2}$
\newauthor and Alexander Wiegand$^{3}$
\\
$^{1}$Center for Astrophysics | Harvard \& Smithsonian, 60 Garden St., MA 02138, USA\\
$^{2}$McWilliams Center for Cosmology, Carnegie Mellon University, Pittsburgh, PA 15213, USA\\
$^{3}$Max-Planck-Institut f{\"u}r Astrophysik, Karl-Schwarzschild-Str. 1, D-85741 Garching, Germany\\
}

\date{Accepted XXX. Received YYY; in original form ZZZ}

\pubyear{2019}

\begin{document}
\label{firstpage}
\pagerange{\pageref{firstpage}--\pageref{lastpage}}
\maketitle

\begin{abstract}
To make use of clustering statistics from large cosmological surveys, accurate and precise covariance matrices are needed. We present a new code to estimate large scale galaxy two-point correlation function (2PCF) covariances in arbitrary survey geometries \resub{that, due to new sampling techniques, runs $\sim 10^4$ times faster than previous codes, computing finely-binned covariance matrices with negligible noise in less than 100 CPU-hours.} As in previous works, non-Gaussianity is \resub{approximated via a small rescaling of shot-noise in the theoretical model, calibrated by comparing jackknife survey covariances to an associated jackknife model. The flexible code, \texttt{RascalC}, has been publicly released, and automatically takes care of all necessary pre- and post-processing, requiring only a single input dataset (without a prior 2PCF model).} Deviations between large scale model covariances from a mock survey and those from a large suite of mocks are found to be be indistinguishable from noise. In addition, the choice of input mock are shown to be irrelevant for desired noise levels below $\sim 10^5$ mocks. Coupled with its generalization to multi-tracer data-sets, this shows the algorithm to be an excellent tool for analysis, reducing the need for large numbers of mock simulations to be computed.
\end{abstract}




\section{Introduction}\label{sec: intro}
With the next generation of cosmological surveys fast approaching, it is of paramount importance to develop formalisms for creating data covariance matrices to estimate uncertainties on derived cosmological parameters. Future surveys will allow unprecedented exploration of cosmic structure formation and Baryonic Acoustic Oscillations (BAO), through upcoming projects such as Euclid \citep{2011arXiv1110.3193L}, DESI \citep{2013arXiv1308.0847L} and WFIRST \citep{2018arXiv180403628D}. With accurate knowledge of experimental covariances, we will additionally be able to perform analysis using cross-correlations of data from different tracer galaxies or surveys.

Conventionally, covariance matrices are obtained from a large suite of `mock' catalogs, simulating large cosmological surveys. Derived two-point correlation functions (2PCFs) for each mock are then combined to estimate the survey covariance. For mocks to be a useful predictor of covariances, we require them to be (a) accurate, such that there is limited bias in the \resub{(two- and higher-point) correlation functions and covariance estimates}, and (b) numerous, to drive down the noise levels in the computed matrices, \resub{degrading final parameter estimates} \citep{2013AJ....145...10D,2014MNRAS.442.2728T,2014MNRAS.439.2531P}. This clearly requires substantial computational power, thus we should look to approximate methods to obtain such covariances in a fraction of the time. In this paper, we build upon the techniques of \citet{2016MNRAS.462.2681O} and \citet{2019MNRAS.487.2701O}, providing a new algorithm to estimate galaxy covariance matrices from only a single survey in a fraction of the previous computational time. In addition, we extend the formalism to compute covariance matrices between \textit{multiple} galaxy tracer populations.

Numerous works have demonstrated the effects of noise in the covariance matrix \citep[e.g.][]{2013MNRAS.432.1928T,2013AJ....145...10D,2014MNRAS.439.2531P}, \resub{showing that noise inflates the width of parameter constraints relative to an ideal measurement by $\mathcal{O}(1/(N_\mathrm{mocks}-N_\mathrm{bins}))$} for a 2PCF estimated in $N_\mathrm{bins}$ using $N_\mathrm{mocks}$. As survey depth increases, so does the number of 2PCF bins, $N_\mathrm{bins}$, particularly with the emerging trend of tomographic analysis in current and future surveys (e.g. \resub{DES; \citealt{2018PhRvD..98d3526A},} eBOSS; \citealt{2019MNRAS.482.3497Z} \resub{and, for weak lensing analyses, KiDS; \citealt{2017MNRAS.465.1454H}}). To obtain the same level of precision in our covariance matrices we hence need more numerous and accurate mocks, leading to a steady increase in computational power required. Since such resources are still in limited supply, it is desirable to search for alternative solutions, in particular approximate methods for covariance matrix generation \resub{This is usually achieved via covariance matrix modeling \citep[e.g.][]{2016MNRAS.457..993P,2016MNRAS.457.1577G,2016MNRAS.462.2681O,2019MNRAS.482.1786L}, numerical techniques \citep[e.g.][]{2015MNRAS.454.4326P,2017MNRAS.466L..83J} or hybrid approaches \citep[e.g.][]{2018MNRAS.473.4150F}, and allows for low-noise covariance matrices to be computed using far fewer mocks than canonical approaches (see  \citet{2019MNRAS.482.1786L} for a comparison of such methods).}

Approximations of the purely Gaussian covariances are provided by \citet{2016MNRAS.457.1577G} and \citet{2019JCAP...01..016L}, where the former is limited to uniform cubic surveys (and hence is not representative of observational data) yet the latter gives an analytic formalism for computing the 2PCF covariance for an arbitrary survey geometry \resub{(building upon the work of \citet{2011MNRAS.414..329C} for the angular 2PCF)}. \resub{\citet{2014MNRAS.440.1379E} and} \citet{2016MNRAS.457..993P} extend this to include non-Gaussianity, with the latter work using a seven-parameter model of the survey-dependent power spectrum covariance that can be calibrated with (a small number of) simulations. Similarly, in \citet{2016MNRAS.462.2681O}, it was shown that, for a single set of tracer galaxies, a configuration-space covariance matrix can be well represented by a theoretical approximation computed solely from 2PCFs, given knowledge of the survey geometry and weight functions. Non-Gaussianity was found to be incorporated effectively by a simple rescaling of the galactic shot-noise, again calibrated via a small number of mock galaxy catalogs. This produced covariance matrix estimates of comparable precision using far fewer mocks than purely mock-based approaches.

\resub{In this paper we will use the shot-noise rescaling introduced in \citet{2016MNRAS.462.2681O} to model non-Gaussian contributions to the covariance matrix. The physical intuition behind this is as follows. Non-Gaussian covariance matrix terms are sourced by higher-point correlation functions, which primarily describe additional clustering on small scales (usually less than the binning width). By enhancing the shot-noise, we increase the clustering power on \textit{infinitely} small scales, which is found to be a good approximation in practice. Mathematically, this is equivalent to replacing integrals over the higher-point correlation functions with their contractions (which are simply shot-noise terms); a fair assumption if the functions are dominated by their squeezed limits on the relevant scales. Additional motivation comes from parallels with Effective Field Theory, with higher order corrections being absorbed into a `renormalized' shot-noise parameter. Whilst this cannot significantly modify clustering at large galaxy separations, this is not found to be an issue since the galaxy field appears to be highly Gaussian on these scales. In \citet{2018MNRAS.477.1153V}, the validity of this approach was clearly demonstrated in its application to BAO parameter constraints, although we expect the approximation to work less well on smaller scales, such as for quasi-linear redshift-space distortions.}

A number of recent papers have considered the determination of 2PCF covariance matrices from only small volumes \resub{(e.g. \citet{2017MNRAS.472.4935H}, \citet{2018MNRAS.478.4602K} and, for super-sample covariance, \citet{2018JCAP...06..015B})}, showing this to be a viable approach to constraining larger scale covariance matrices. This was further developed in \citet{2019MNRAS.487.2701O}, where the technique of constraining the shot-noise rescaling directly from the data was introduced, splitting the data into small \textit{jackknife regions} and computing the 2PCFs in each. Given these estimates, a jackknife covariance matrix was obtained, which was compared to a theoretical prediction in order to compute the shot-noise rescaling. This approach thus allows covariance matrices to be estimated purely from data, without calibration with mocks, and additionally allows us to use any input survey geometry.

In this paper, we introduce an improved \resub{sampling} algorithm, which is able to produce low-noise estimates of the covariance matrix in a fraction of the previous computation time. We adopt a similar \resub{jackknife-calibrated} rescaling approach to \citet{2019MNRAS.487.2701O}, yet switch to the \textit{unrestricted} jackknife formalism, which allows for more precise determinations of rescaling parameter. Covariance matrix computation is greatly expedited by a new grid-based importance sampling technique, drawing sampling points from random galaxy position catalogs (such as those used for $RR$ pair counts) rather than newly generating them from some mask and number density distribution. \resub{The updated sampling allows the code to run $\sim 10^4$ times faster than previous codes \citep{2016MNRAS.462.2681O} with no loss of accuracy.}

With new data releases from galaxy surveys such as the Extended Baryon Oscillation Spectroscopic Survey \citep[eBOSS; ][]{2016AJ....151...44D}, we will soon be able to obtain improved constraints on large scale structure parameters by using multiple galaxy tracers, for example the Emission Line Galaxies (ELG) and Luminous Red Galaxies (LRG) populations. For such analyses, we require \textit{cross-covariance} matrices, evaluating the associations between auto- and cross-correlation functions. The addition of such complexity conventionally requires more mocks to be computed; yet via our approach, arbitrary cross-covariances can be computed with ease. We describe an extended formalism to compute these matrices from data alone, with all possible two-tracer covariance matrices requiring only $\sim 6$ times the computation time of a single tracer survey.

The full analysis pipeline for this fast and flexible approach has been condensed into a publicly available C\texttt{++} and Python package; \texttt{RascalC}\footnote{\url{https://github.com/oliverphilcox/RascalC}} with extensive documentation showing its usage for single- and multiple-tracer cosmology.\footnote{\url{https://RascalC.readthedocs.io}} Included are all relevant routines allowing estimation of a set of covariance matrices simply from input survey or random galaxy position files in sky or Cartesian coordinates. The flexibility of the code ensures that it can be simply altered to take more complex forms for the theoretical covariance matrix, for example with some estimation of the 3- or 4-point correlation functions, or a different jackknife formalism. The authors are happy to assist with this process on request.

The structure of this paper is as follows. In Sec.\,\ref{sec: theory} we present the theoretical covariance matrices for the new unrestricted jackknife formalism, with their evaluation discussed in Sec.\,\ref{sec: matrix_evaluation}, \resub{which outlines our new (and highly efficient) sampling scheme}. Sec.\,\ref{sec: algorithm_design} gives a description of the computational algorithm used here, followed by a comparison to previous code and an application to simulated Quick Particle Mesh \citep[QPM;][]{2014MNRAS.437.2594W} data in Sec.\,\ref{sec: results}. The approach is generalized to deal with multiple galaxy tracer populations in Sec.\,\ref{sec: general_cov_matrices} before we finish with conclusions and mathematical derivations in Sec.\,\ref{sec: conclusions} and appendices \ref{appen: jackknife_expansion}-\ref{appen: KL-div}.

\section{Theoretical Covariance Matrix Estimators}\label{sec: theory}
We begin by considering the covariance matrix estimators used in this paper. For clarity, we first recapitulate the full covariance matrix integrals given in \citet{2016MNRAS.462.2681O} before discussing the modification to jackknife covariance matrices. \resub{Throughout this section, we assume space to be discretized into a set of small cells (each of which can contain at most one galaxy) to simplify our expressions. As in \citet{2016MNRAS.462.2681O} and \citet{2019MNRAS.487.2701O} the indices $i,j,k,l$ refer to these cells and not the galaxies found therein.}

\subsection{Full Covariance Matrices}
The standard estimator for the anisotropic 2PCF is given by
\beq\label{eq: single_full_xi}
    \hat{\xi}_{a} = \frac{1}{RR_{a}}\sum_{i\neq{}j}D_a^{ij}, \qquad
    RR_{a} = \sum_{i\neq{}j}R_a^{ij} 
\eeq
\citep{1993ApJ...412...64L}, \resub{summing over all pairs of cells (denoted $i$ and $j$)} and using the definitions
\beq
    D_a^{ij}&=&\Theta_a^{ij}n_in_jw_iw_j\delta_i\delta_j, \qquad
    R_a^{ij}=\Theta_a^{ij}n_in_jw_iw_j
\eeq
where $n_i$ and $w_i$ are the mean number density and weight in cell $i$ and $\Theta_a^{ij}$ is a binning function (which is unity if the pair of cells lies in the bin $a$ and zero else). Here, the observed galaxy density is $n_i^{(g)} = n_i(1+\delta_i)$, such that $\av{n_i^{(g)}}=n_i$, and the weights are set by the utilized survey, e.g. FKP weights \citep{1994ApJ...426...23F} in mock catalogs for the Baryon Oscillation Spectroscopic Survey \citep[BOSS;][]{2015ApJS..219...12A,2017MNRAS.470.2617A}. From $\hat\xi_a$ we may construct the covariance matrix via
\beq
    \hat{C}_{ab} = \av{\hat\xi_a\hat\xi_b}-\av{\hat\xi_a}\av{\hat\xi_b}=  \frac{1}{RR_aRR_b}\sum_{i\neq j}\sum_{k\neq l}\left[\av{D^{ij}_aD^{kl}_b}-\av{D^{ij}_a}\av{D^{kl}_b}\right].
\eeq
In \citet{2016MNRAS.462.2681O} it was shown that, by expanding the summation into 2-, 3- and 4-point terms and using Wick's theorem to expand $\av{\delta_i...\delta_j}$ factors, this leads to the full survey theoretical covariance estimator in bins $a,b$
\beq\label{eq: C_full_form_no_alpha}
    \av{C_{ab}} = C_{4,ab}+C_{3,ab}+C_{2,ab},
\eeq
with components defined by
\beq\label{eq: C234FullIntegrals}
    C_{4,ab} &=& \frac{1}{RR_aRR_b}\sum_{i\neq j\neq k\neq l}n_in_jn_kn_lw_iw_jw_kw_l\Theta_a^{ij}\Theta_b^{kl}\left[\xi^{(4)}_{ijkl}+2\xi_{ik}\xi_{jl}\right]\\\nonumber
    C_{3,ab} &=& \frac{4}{RR_aRR_b}\sum_{i\neq j\neq k}n_in_jn_kw_iw_j^2w_k\Theta_a^{ij}\Theta_b^{jk}\left[\xi^{(3)}_{ijk}+\xi_{ik}\right]\\\nonumber
    C_{2,ab} &=& \frac{2\delta_{ab}}{RR_aRR_b}\sum_{i\neq j}n_in_jw_i^2w_j^2\Theta_a^{ij}\left[1+\xi_{ij}\right].
\eeq
\resub{Here, we denote the 2PCF by $\xi_{ij} \equiv \xi(\vec r_i-\vec r_j)$ (for cell centers at $\vec r_i$ and $\vec r_j$) and the three- (four-)point correlation function by $\xi^{(3)}$ ($\xi^{(4)}$).} Note that we have exploited the relabelling symmetry of the 3-point matrix to exchange $i$ and $j$ with respect to \citet[Eq.\,2.25]{2016MNRAS.462.2681O}; this is done for later efficiency. \resub{These may be alternatively written in continuous space via the replacements $\sum_i\rightarrow\int d^3\vec r_i$, $X_i\rightarrow X(\vec r_i)$ and $X_{ij}\rightarrow X(\vec r_i - \vec r_j)$.}

An important assumption in this paper (and in previous works) is that non-Gaussianity can be well approximated by simply rescaling the level of shot-noise in the survey by a factor $\alpha$,\footnote{For clarity, we denote the shot-noise parameter as $\alpha$ rather than the previously-used $a$.} \resub{as justified in the introduction.} This can be calibrated either by jackknives or mock data, and allows us to drop the higher point correlation functions $\xi^{(3)}$ and $\xi^{(4)}$ in the above expressions giving the new estimator
\beq\label{eq: C_full_form}
    \av{C_{ab}} = C_{4,ab}+\alpha C_{3,ab}+\alpha^2 C_{2,ab}.
\eeq

\subsection{Jackknife Covariance Matrices}
\subsubsection{The Unrestricted Jackknife Formalism}
One of the key benefits of a large cosmological survey is the ability to split the survey into sub-regions to produce multiple estimates of quantities such galaxy correlation functions. A standard approach to this is to use \textit{jackknives}, where we split the survey into $N_J$ regions (as depicted in Fig.\,\ref{fig: cell-cartoon}), and compute an estimate of the 2PCF with each region left out in turn \citep[e.g.][]{2009MNRAS.396...19N,2016MNRAS.456.2662F}. \resub{In this work, the jackknife covariance matrices are \textit{not} used as estimates of the full-volume covariance matrix, but rather to constrain the shot-noise rescaling parameter $\alpha$. This is done by comparing theoretical and data-derived jackknife covariance matrices, and the optimal $\alpha$ is then passed to the full covariance estimator (Eq.\,\ref{eq: C_full_form}), giving an estimate of the (non-Gaussian) complete survey covariance.}

As noted in \citet{2019MNRAS.487.2701O}, cosmological jackknives suffer from additional complexities not found in traditional jackknife approaches, most notably that different jackknife regions are not independent and pairs of \resub{cells} (containing galaxy positions from the random or survey catalogs) can straddle jackknife regions, which is important for 2PCF analyses, where we are primarily interested in pair counting. There exist several choices of jackknife formalism corresponding to different weightings between \resub{cells (or particles in the continuous limit)}, denoted $q_{ij}^A$ for \resub{cells} $i$ and $j$ and jackknife region $A$. For the \textit{restricted jackknife}, as used in \citet{2019MNRAS.487.2701O}, we assign unit weight to pairs of \resub{cells} that both lie in region $A$ and zero weight else, giving $q_{ij}^A = q_i^Aq_j^A$, where $q_i^A$ is unity if $i$ is in region $A$ and zero else. In this paper, we instead adopt the \textit{unrestricted jackknife} formalism, which includes \resub{cell} pairs which straddle the jackknife region. We assign half-weight to \resub{cell} pairs where only one \resub{cell} is in the region and unit weight to pairs which are both in the same region, giving the modified form 
\beq\label{eq: unrestricted_jk_weights}
    q_{ij}^A = \frac{1}{2}\left(q_i^A+q_j^A\right), \qquad    Q_{ij} \equiv \sum_A q_{ij}^A = 1.
\eeq
This allows us to include all \resub{cell} pairs and to probe the 2PCF on scales larger than the jackknife region. \resub{Whilst these jackknifes are by no means independent, this is not an issue in our analysis because the jackknife covariance is not being used to estimate the full-volume covariance directly. (This avoids having to introduce additional complexities such as the `pair jackknife' of \citet{2016MNRAS.456.2662F}, which assigns half of all particle pairs crossing a given jackknife region to the first region and half to the second to remove bias.)} Since pair counts are by nature additive, it is true here that the covariance matrix of jackknife correlation functions can be rewritten as a rescaled version of the \textit{sample} covariance between the estimates of the 2PCF $\xi$ \textit{in} each jackknife region (rather than those \textit{excluding} each region). This simplifies the analysis greatly, and hereafter we will only consider estimates of $\xi$ within each region. 

\subsubsection{Correlation Function Estimators}
In a jackknife region $A$, the 2PCF may be estimated \resub{as a sum over fine cells, $i$, $j$,} in an analogous fashion to Eq.\,\ref{eq: single_full_xi}:
\beq\label{eq: single_jackknife_xi}
    \hat{\xi}_{aA} = \frac{1}{RR_{aA}}\sum_{i\neq{}j}q_{ij}^A D_a^{ij}, \qquad    RR_{aA} = \sum_{i\neq{}j}q_{ij}^AR_a^{ij} 
\eeq
for arbitrary jackknife weighting function $q_{ij}^A$. From these, an overall estimate is defined via
\beq\label{eq: xiJ_estimator}
    \hat{\xi}^J_a = \sum_{A}w_{aA}\hat{\xi}_{aA} = \sum_A \frac{RR_{aA}}{RR_a^J}\hat\xi_{aA}
\eeq
where $RR_a^J = \sum_A RR_{aA}$. The weights, $w_{aA} = \frac{RR_{aA}}{RR_a^J}$, are analogous to the volume fraction of each region, but allow for variations between 2PCF bins. Utilizing the definitions of Eq.\,\ref{eq: single_jackknife_xi}, this becomes
\beq\label{eq: xiJ_estimator_expanded}
    \hat{\xi}^J_a = \frac{1}{RR_{a}^J}\sum_{i\neq{}j}Q_{ij}D_a^{ij}, \qquad
    RR_a^J = \sum_{i\neq{}j}Q_{ij}R_a^{ij}
\eeq
for $Q_{ij} = \sum_Aq_{ij}^A$. For the unrestricted jackknife, where $Q_{ij} \equiv 1$, $RR_a^J$ and $\hat\xi_a^J$ reduce to $RR_a$ and $\hat\xi_a$; the full survey forms (cf.\,Eq.\,\ref{eq: single_full_xi}). 

\subsubsection{Covariance Matrix Integrals}
As in \citet{2019MNRAS.487.2701O}, we can compute the weighted jackknife covariance matrix (in bins $a,b$) from the jackknife 2PCFs as 
\beq\label{eq: C_jackknife_estimator} 
\hat{C}_{ab}^J = \frac{1}{1-\sum_B\bar{w}_{aB}\bar{w}_{bB}}\left[\sum_A{\bar{w}_{aA}\bar{w}_{bA}\left(\hat{\xi}_{aA}-\hat{\xi}_a^J\right)\left(\hat{\xi}_{bA}-\hat{\xi}_b^J\right)}\right],
\eeq
which can be used both to construct a theoretical model of the covariance matrix and to find an estimate directly from data, using $\hat\xi_{aA}$ estimates from exhaustive pair counting. The weighting function $\bar{w}_{aA}$ is here set equal to the 2PCF weighting $w_{aA}$.\footnote{The exact choice of $\bar w_{aA}$ is not important here, since we attempt only to fit a theoretical estimate of $\hat C_{ab}^J$ to one constructed from data.} 

Since each unrestricted jackknife uses all particles in the survey, we expect non-negligible correlations between individual $\hat\xi_{aA}$ estimates (expected to be negative for a uniform survey), meaning that they are not strictly independent. This implies that the jackknife covariance matrix $\hat C_{ab}^J$ is \textit{not} a good estimator of the full-survey covariance, and is likely to be an underestimate on average. In our context, the jackknife covariance matrix is being used solely to fit for the shot-noise rescaling parameter $\alpha$, rather than estimate the full-survey covariance and, since this effect is common to both theory and data $\hat C^J_{ab}$ estimates, it may be ignored.

Eq.\,\ref{eq: C_jackknife_estimator} may be expanded analogously to the full covariance matrix estimator, and a full derivation is presented in Appendix \ref{appen: jackknife_expansion}. As before, we assume a shot-noise rescaling parameter $\alpha$ and expand the summations and random field expectations to yield the estimator\footnote{\resub{It is not obvious \textit{a priori} that the jackknife covariance will have the same shot-noise rescaling parameter $\alpha$ as the full-covariance matrix, as the two matrices are different in expectation. In \citet{2019MNRAS.487.2701O}, the approach was tested using the restricted jackknife formalism and estimates of $\alpha$ from jackknifes were found to agree with those from mocks at high precision.}}
\beq\label{eq: NewFullJackknifeC2-4Expression}
    \av{C^J_{ab}} = C^J_{4,ab}+\alpha{}C^J_{3,ab}+\alpha^2C^J_{2,ab},\eeq
where we define
\beq\label{eq: NewC234Definitions} 
    C^J_{4,ab} &=&  \frac{1}{RR_a^JRR_b^J-\sum_BRR_{aB}RR_{bB}}\sum_{i\neq{}j\neq{}k\neq{}l}n_in_jn_kn_lw_iw_jw_kw_l\Theta^{ij}_a\Theta^{kl}_b\left(\xi^{(4)}_{ijkl}+\xi_{ij}\xi_{kl}+2\xi_{ik}\xi_{jl}\right)\omega_{ijkl}^{ab}\\\nonumber
    C^J_{3,ab} &=&  \frac{4}{RR_a^JRR_b^J-\sum_BRR_{aB}RR_{bB}}\sum_{i\neq{}j\neq{}k}n_in_jn_kw_iw_j^2w_k\Theta^{ij}_a\Theta^{jk}_b\left(\xi^{(3)}_{ijk}+\xi_{ik}\right)\omega_{ijjk}^{ab}\\\nonumber
    C^J_{2,ab} &=&  \frac{2\delta_{ab}}{RR_a^JRR_b^J-\sum_BRR_{aB}RR_{bB}}\sum_{i\neq{}j}n_in_jw_i^2w_j^2\Theta^{ij}_a\left(1+\xi_{ij}\right)\omega_{ijij}^{ab}
\eeq
(including the non-Gaussian terms for generality, although these are not used in the computation). The weighting tensor $\omega_{ijkl}^{ab}$ is given by
\beq\label{eq: omega_tensor_definition}
\omega_{ijkl}^{ab} = \sum_A\left[\left(q_{ij}^A-w_{aA}Q_{ij}\right)\left(q_{kl}^A-w_{bA}Q_{kl}\right)\right].
\eeq
and, in the case of the unrestricted jackknife, simplifies to 
\beq\label{eq: omega_tensor_expansion}
    \omega_{ijkl}^{ab} &=& \frac{1}{4} \sum_A\left(q_{i}^{A}q_{k}^{A}+q_{j}^{A}q_{k}^{A}+q_{i}^{A}q_{l}^{A}+q_{j}^{A}q_{l}^{A}\right)- \frac{1}{2}\left(w_{bJ_i}+w_{bJ_j}+w_{aJ_k}+w_{aJ_l}\right)+ \sum_{A}w_{aA}w_{bA}.
\eeq
via Eq.\,\ref{eq: unrestricted_jk_weights}, where we define $J_x$ as the jackknife region of cell $x$. \resub{Equivalent expressions for the jackknife covariance matrix terms in continuous space are given in Eqs.\,\ref{eq: C234JackIntegrals} in the appendix.} \resub{The theoretical jackknife covariance model defined by Eqs.\,\ref{eq: NewC234Definitions} will later be fit to the sample jackknife covariance of Eq.\,\ref{eq: C_jackknife_estimator} (using measured 2PCF estimates in each jackknife) to constrain the shot-noise rescaling parameter $\alpha$.}

We note significant differences between the $\av{C_{ab}^J}$ forms of \citet{2019MNRAS.487.2701O} and those here. In the former, the jackknife covariance was taken as $\av{\hat\xi_a^J\hat\xi_b^J}-\av{\hat\xi_a^J}\av{\hat\xi_b^J}$; we instead use the full jackknife covariance of Eq.\,\ref{eq: C_jackknife_estimator} for better comparison with observational data, which gives a different prefactor, an additional $\xi_{ij}\xi_{kl}$ term (cf.\,Sec.\,\ref{subsec: disconnected_term}) and new $\omega_{ijkl}^{ab}$ weighting tensors. 

\subsubsection{The Disconnected Term}\label{subsec: disconnected_term}
In the expectation of the $C_{ab}^J$ integral, we note the presence of a \textit{disconnected} 4-point term involving $\xi_{ij}\xi_{kl}$ which may be factored into a product of 2-point terms, summed over the jackknife region $A$. This was not present in \citet{2019MNRAS.487.2701O} and arises due to the different form of the jackknife covariance matrix adopted. Returning to the summation form of the integrals, we may rewrite this term (hereafter denoted $C_{x,ab}^J$) in the following separable form:\footnote{Note that we switch from $i\neq j\neq k\neq l$ summations to a pair of $i\neq j$ and $k\neq l$ summations here. Whilst this generates two additional lower-point terms, both involve an additional factor of $n(\vec r)$ in the integrand (unlike for the $\delta_i\delta_i$ contractions in the main integrals), giving negligible contributions to: the integrated term.}
\beq\label{eq: disconnected_term}
C_{x,ab}^J &=& \frac{1}{RR_a^JRR_b^J-\sum_BRR_{aB}RR_{bB}}\sum_{i\neq j\neq k\neq l}\left(n_in_jw_iw_j\Theta_a^{ij}\xi_{ij}\right)\left(n_kn_lw_kw_l\Theta_b^{kl}\xi_{kl}\right)\omega_{ijkl}^{ab}\\\nonumber
&=& \frac{1}{RR_a^JRR_b^J-\sum_BRR_{aB}RR_{bB}}\sum_A\left[EE_{aA}-w_{aA}EE_a^{J}\right]\left[EE_{bA}-w_{bA}EE_b^{J}\right]
\eeq
where we define 
\beq\label{eq: EEaA definition}
EE_{aA} &=& \sum_{i\neq{}j}q_{ij}^An_in_jw_iw_j\Theta_a^{ij}\xi_{ij}, \qquad EE_a^J = \sum_AEE_{aA} = \sum_{i\neq{}j}Q_{ij}n_in_jw_iw_j\Theta_a^{ij}\xi_{ij}
\eeq
analogously to $RR_{aA}$. In realistic surveys, the disconnected term is expected to be small, and the conditions under which it cancels entirely are discussed in appendix \ref{appen: disconnected}.

\begin{figure}%
    \centering
    \includegraphics[width=0.7\textwidth]{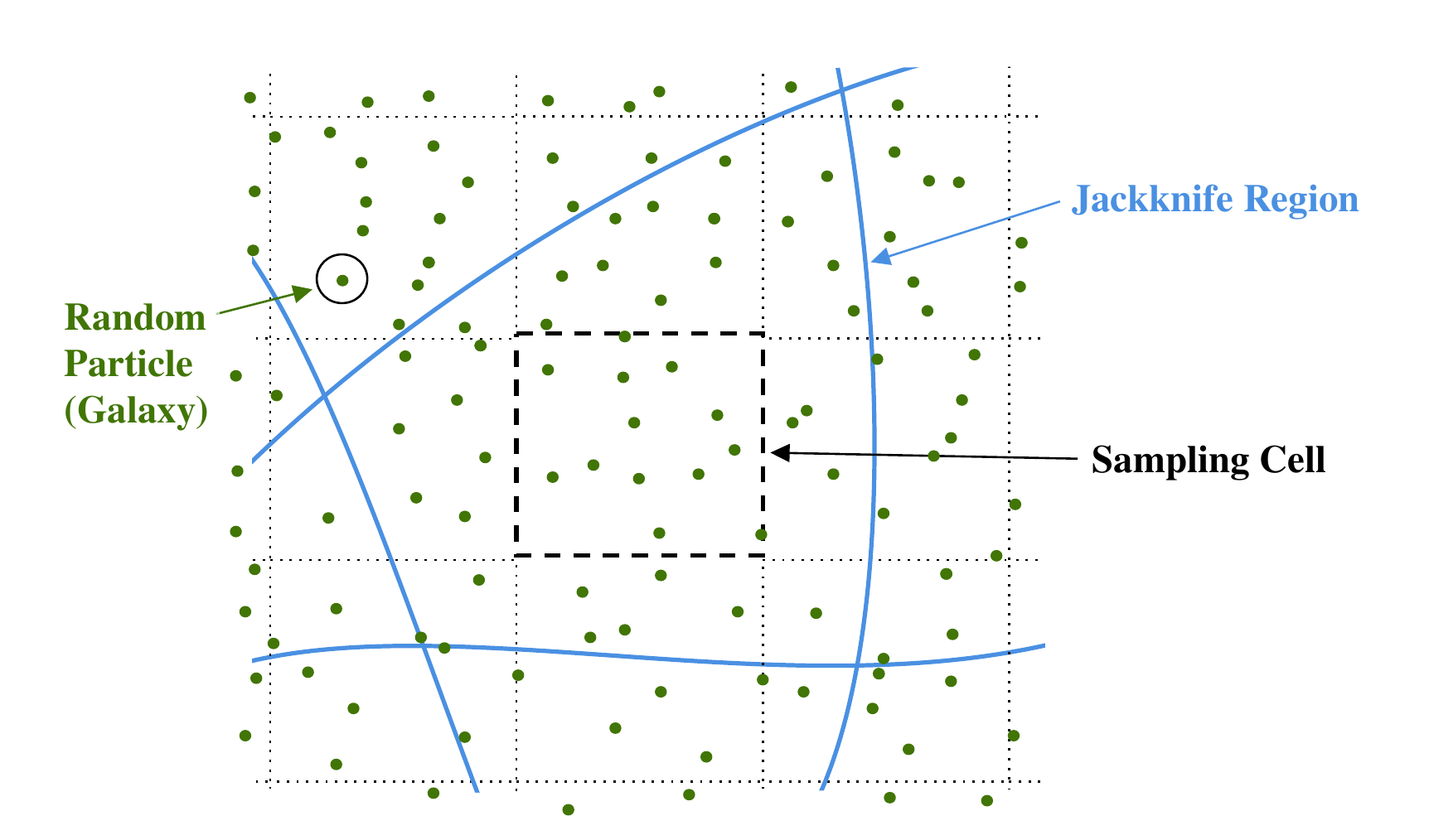}%
    \caption{Cartoon of a small 2D slice in configuration space illustrating the various survey subdivisions used in this paper. The green points represent `random particles'; galaxies drawn from a survey catalog of `randoms' with number density following that of the survey selection function. These are partitioned into contiguous jackknife regions, the boundaries of which are depicted by full blue lines (defined by \texttt{HEALPix} pixels in this analysis). The dotted lines indicate the boundaries of 3D `sampling cells', together forming the `sampling grid'. These are used to give efficient Monte Carlo sampling of the covariance integrals, and typically have side lengths $\sim 10\,\mathrm{Mpc}/h$, containing $\lesssim 10$ random particles per sampling cell. The jackknife regions used in this paper have diameter $\sim 200\,\mathrm{Mpc}/h$ thus are not to scale here.}%
    \label{fig: cell-cartoon}%
\end{figure}

\section{Evaluation of the Covariance Matrix Integrals}\label{sec: matrix_evaluation}

\resub{Computation of the covariance matrix integrals derived above (Eqs.\,\ref{eq: C234FullIntegrals} and \ref{eq: NewC234Definitions}) allows us to compute a useful full-volume model covariance matrix. To evaluate these terms, we will make use of a new sampling strategy, which is described below. This divides up the survey into `sampling cells' (formalized in Sec.\,\ref{subsec: grids}) and uses Monte Carlo importance sampling methods to pick these in an efficient manner, giving an algorithm that is faster than previous codes \citep{2016MNRAS.462.2681O} by a substantial factor ($\sim 10^4$).}

\subsection{The Sampling Grid Formalism}\label{subsec: grids}
To efficiently compute the high-dimensional integrals in Eqs.\,\ref{eq: C234FullIntegrals}\,\&\,\ref{eq: NewC234Definitions}, we require Monte Carlo sampling methods, where we draw sets of $m\in\{2,3,4\}$ points in three-dimensional survey space (hereafter `pairs', `triples' and `quads') and use these to evaluate the $3m$-dimensional integrals via summation of the integrands \resub{(see \citet{2004A&A...413..465K} and \citet{2008A&A...477...43J} for an application of similar ideas to cosmic shear covariance in configuration- and Fourier-space respectively).} In previous work \citep{2016MNRAS.462.2681O,2019MNRAS.487.2701O} using the \texttt{Rascal} code,\footnote{\url{https://github.com/rcoconnell/Rascal}} random sampling positions were generated from the survey geometry defined by the experiment mask and an estimate of the redshift-dependent number density $n(\vec r)$ and particle weight $w(\vec r)$. Here, we adopt a new approach, making no use explicit use of the mask, but instead using catalogs of `randoms' provided by galaxy surveys, consisting of random galaxy positions (denoted `random particles') distributed according to the survey selection function. These are primarily used to compute $RR$ counts for 2PCF computations. Their number densities, $n^r(\vec r)$ are expected to have the same mean spatial dependence as for the survey galaxy distribution, $n(\vec r)$,\footnote{When this approach is applied to the jackknives, there is an implicit assumption that the galaxy-to-random-particle ratio $N_g/N_r$ is constant between jackknives. Our formalism could be generalized to account for this, but, since jackknives are used only for calibration, the problem can be avoided by fixing the jackknife pair counts normalization to the full-survey $N_g/N_r$ value, as in Sec.\,\ref{subsec: corr_functions}.} and we assume the distribution of random particle weights matches that of the galaxies.\footnote{It may seem tempting to include the weights in the normalization to allow for differences between $w(\vec r)$ and $w^r(\vec r)$; this is not helpful in this context, since the 2- and 3-point integrals differ in the number of $n(\vec r)$ and $w(\vec r)$ factors.}  We may thus write
\beq\label{eq: ngal-nrand}
    n(\vec r)= \frac{N_g}{N_r}n^r(\vec r) = \frac{\int \mathrm{d}^3\vec r\,n(\vec r)}{\int \mathrm{d}^3\vec r\,n^r(\vec r)}n^r(\vec r), \qquad 
    w(\vec r)= w^r(\vec r)
\eeq 
for a total of $N_g$ and $N_r$ galaxies and random particles respectively.

For computational efficiency, these random particles are placed on a cuboidal \textit{sampling grid} encompassing the entire survey region, partitioning them into a number of cubic \textit{sampling cells}, each of which contains a small number $\lesssim10$ particles, as illustrated in Fig.\,\ref{fig: cell-cartoon}. (These cells are distinct from the small regions used to define the discrete summations in Sec.\,\ref{sec: theory} which contained at most one particle.) This sampling grid-based approach allows for fast and trivially parallelizable code, where we can pre-assign probabilities of picking each sampling cell according to some given sampling strategy, ensuring that all elements of the covariance matrices are well sampled. In this approach, there is no explicit dependence on a survey mask which is useful since (a) future surveys may only generate random particles rather than a mask lookup function and (b) there is a large computational efficiency boost, as sampling reduces to picking random sampling cells, without having to query a complex mask.

\subsection{Importance Sampling Techniques}
To evaluate high-dimensional integrals, Monte Carlo techniques are required, in particular \textit{importance sampling}, where we pick points in parameter-space in such a way as to allow for fast convergence of the integral. To see how this may be applied in practice, first consider the integral of an arbitrary function of a single (3D) spatial coordinate $X(\vec r)$ over the survey;
\beq
    I = \int d^3\vec r \,f(\vec r)X(\vec r)
\eeq
where $f(\vec r)$ is some normalized weighting function satisfying $\int d^3\vec r \,f(\vec r) = 1$. Treating $f(\vec r)$ as a probability density function (PDF) for $\vec r$, this may be rewritten as an expectation over $f$, $\mathbb{E}_f[X]$, which can be sampled by drawing random $\vec r_i$ positions according to the PDF $f$;
\beq
    I = \mathbb{E}_f[X] \approx \frac{1}{N_\mathrm{draws}}\sum_{\vec r_i\sim f}X(\vec r_i)
\eeq
where we make a total of $N_\mathrm{draws}$ from $f$ (the approximation becomes exact in the limit $N_\mathrm{draws}\rightarrow\infty$). Importance sampling tells us that we may also draw points from a different PDF, $g(\vec r)$, by writing
\beq
    I = \int d^3\vec r\,g(\vec r)X(\vec r)\frac{f(\vec r)}{g(\vec r)} = \mathbb{E}_g\left[X\frac{f}{g}\right] \approx \frac{1}{N_\mathrm{draws}}\sum_{\vec r_i\sim g}X(\vec r_i)\frac{f(\vec r_i)}{g(\vec r_i)}
\eeq
where the expectation is now over $g$ and positions are sampled from this PDF, with a reweighting by ${f(\vec r)}/{g(\vec r)}$. Whilst this may not seem useful, it allows us to manually choose the sampling distribution $g$ and alter the estimator's efficiency. The optimal function $g^*(\vec r)$ is found by minimizing the variance of the estimator for $I$; this can be shown to be proportional to $X(\vec r)f(\vec r)$. In practice, we cannot sample directly from $g^*(\vec r)$, thus we instead choose some $g(\vec r)$ which is (a) easy to sample from and (b) similar in form to $X(\vec r)f(\vec r)$. This allows us to preferentially sample regions where the summand is large, leading to faster convergence.

In this paper, we are interested in sampling integrals which depend on sets of two, three and four points in space. To demonstrate the extension of the above methodology to higher dimensional cases, we consider a two-point case; the integral of a function $X(\vec r_i,\vec r_j)$ weighted by the product of (continuous) \textit{galaxy} number densities $n(\vec r)$:
\beq
    I = \int d^3\vec{r}_id^3\vec{r}_jn(\vec{r}_i)n(\vec{r}_j)X(\vec{r}_i,\vec{r}_j) = \left(\frac{N_g}{N_r}\right)^2\int d^3\vec r_i d^3\vec r_j n^r(\vec r_i)n^r(\vec r_j)X(\vec r_i,\vec r_j)
\eeq
converting to random particle densities via Eq.\,\ref{eq: ngal-nrand}. In this case, the normalized PDF for drawing a pair of (ordered) points at $\vec r_i$ and $\vec r_j$ is 
\beq\label{eq: random_pdf}
    f_{ij} = f(\vec{r}_i,\vec{r}_j) = \frac{n^r(\vec{r}_i)n^r(\vec{r}_j)}{N_r^2},
\eeq
which allows us to rewrite the integral as an expectation over $f_{ij}$ and hence a sum (as for the 1-point case above);
\beq\label{eq: continuous_integral_approx}
    I &=& \left(\frac{N_g}{N_r}\right)^2\int{}d^3\vec{r}_id^3\vec{r}_j\,\left[N_r^2f(\vec{r}_i,\vec{r}_j)\right]X(\vec{r}_i,\vec{r}_j) = N_g^2\mathbb{E}_{f}\left[X_{ij}\right]\approx\frac{N_g^2}{N_\mathrm{pairs}}\sum_{(\vec r_i,\vec r_j)\sim{f}}X_{ij}
\eeq
for $X_{ij} = X(\vec r_i,\vec r_j)$, where we draw a total of $N_\mathrm{pairs}$ sampling points. Notably, in the above integral and estimator, there is no inclusion of $n(\vec r)$ or $n^r(\vec r)$ since we are drawing points from a distribution which matches that of the galaxies (unlike in \citealt{2016MNRAS.462.2681O} and \citealt{2019MNRAS.487.2701O}). As before, this may be recast into an expectation over some different PDF $g(\vec r_i,\vec r_j)$ via
\beq
    I = N_g^2\mathbb{E}_g\left[X_{ij}\frac{f_{ij}}{g_{ij}}\right] \approx \frac{N_g^2}{N_\mathrm{pairs}}\sum_{(\vec r_i,\vec r_j)\sim g}X_{ij}\frac{f_{ij}}{g_{ij}}
\eeq

In our context, instead of choosing random pairs of \textit{positions} from the survey geometry defined by some mask, we simply pick pairs of \textit{particles} from the set of all possible pairs of input random galaxies (with $N_r^2$ total pairs). \resub{Henceforth, we will use the roman indices $i$, $j$, etc. to label such particles.} This implies that we must switch to discrete statistics, and the aforementioned PDF (Eq.\,\ref{eq: random_pdf}) is transformed into a uniform probability mass function (PMF) $f_{ij}=1/N_r^2$ (as we draw two ordered particles $i$ and $j$ uniformly from the set of $N_r$ particles). $g_{ij}$ now becomes the PMF for drawing two points from the set of random particles, with some user-defined weighting. Thus, inserting the forms of the PMF, the general 2-point integral $I$ becomes
\beq
I = N_g^2\mathbb{E}_g\left[X_{ij}\frac{f_{ij}}{g_{ij}}\right] = \left(\frac{N_g}{N_r}\right)^2\mathbb{E}_g\left[\frac{X_{ij}}{g_{ij}}\right] = \frac{1}{N_\mathrm{pairs}}\left(\frac{N_g}{N_r}\right)^2\sum_{(i,j)\sim{}g}\left[\frac{X_{ij}}{g_{ij}}\right]
\eeq
where we draw $N_\mathrm{pairs}$ particles according to the PMF $g_{ij}$. This extends naturally to summations of triples and quads of particles, using the appropriate normalization by the number of samples and $\left({N_g}/{N_r}\right)^d$ for the $d$-point integral.

\subsection{Sampling Cell Selection \& Integral Estimators}
\subsubsection{Cell-Based Stochastic Estimators}
It remains to decide how to choose sets of particles \resub{(denoted $i,j,k,l$)} to sample, and hence the PMF, $g^{(d)}$, for the $d$-point integral. In our implementation, this is assisted by the use of the aforementioned sampling grid cells \resub{(denoted $c_i$, $c_j$, $c_k$, $c_l$)}. Given some initial particle $i$ in sampling cell $c_i$, subsequent particles are chosen by first picking a sampling cell $c_j$ via some PMF $p_j$ (depending only on the relative distances between $c_i$ and $c_j$) and choosing one of the particles, $j$, inside at random (from a total of $m_j$ particles in $c_j$). For pairs of random particles $(i,j)$, we obtain a distribution function $g^{(2)}_{ij} \propto {p_j}/{m_j}$, depending only on the occupation and separation of sampling cells $c_i$ and $c_j$. For normalization, we include a prefactor $1/N_r$, which corresponds to the number of possible choices of the $i$ particle. The final form of the general pair estimator for $N_\mathrm{pairs}$ samples is thus
\beq
    \hat{I} = \frac{1}{N_\mathrm{pairs}}\left(\frac{N_g}{N_r}\right)^2\sum_{(i,j)\sim{}g^{(2)}}\left[N_r\frac{m_j}{p_j}\right]X_{ij}
\eeq
which again naturally extends to higher order integrals, e.g. sampling from the probability distribution $g^{(4)}_{ijkl} = (p_jp_kp_l)/(N_rm_jm_km_l)$ for the 4-point term. Notably, the particle-selection PMFs are reduced to simply probability distributions over sampling cell positions, which can be pre-computed efficiently. Using this approach, estimators for the jackknife covariance matrix integrals (Eqs.\,\ref{eq: NewC234Definitions}, assuming Gaussianity) become
\beq\label{eq: normalized_jackknife_estimators}
    \hat{C}^J_{4,ab} &=& \frac{1}{N_\mathrm{quads}}\left(\frac{N_g}{N_r}\right)^4\frac{2}{RR_aRR_b}\frac{1}{1-\sum_Aw_{aA}w_{bA}}\sum_{(i,j,k,l)\sim{}g^{(4)}}\left[N_r\frac{m_jm_km_l}{p_jp_kp_l}\right]w^r_iw^r_jw^r_kw^r_l\Theta_a^{ij}\Theta_b^{kl}\xi_{ik}\xi_{jl}\omega_{ijkl}^{ab}\\\nonumber
    \hat{C}^J_{3,ab} &=& \frac{1}{N_\mathrm{triples}}\left(\frac{N_g}{N_r}\right)^3\frac{4}{RR_aRR_b}\frac{1}{1-\sum_Aw_{aA}w_{bA}}\sum_{(i,j,k)\sim{}g^{(3)}}\left[N_r\frac{m_jm_k}{p_jp_k}\right]w^r_i\left(w^r_j\right)^2w^r_k\Theta_a^{ij}\Theta_b^{jk}\xi_{ik}\omega_{ijjk}^{ab}\\\nonumber
    \hat{C}^J_{2,ab} &=& \frac{1}{N_\mathrm{pairs}}\left(\frac{N_g}{N_r}\right)^2\frac{2\delta_{ab}}{RR_aRR_b}\frac{1}{1-\sum_Aw_{aA}w_{bA}}\sum_{(i,j)\sim{}g^{(2)}}\left[N_r\frac{m_j}{p_j}\right]\left(w^r_iw^r_j\right)^2\Theta_a^{ij}\left(1+\xi_{ij}\right)\omega_{ijij}^{ab},
\eeq
\resub{where the summations run over individual \textit{random particle positions}.} (See Sec.\,\ref{subsec: disconnected_evaluation} for the disconnected term estimator). This gives the correct normalization factors to be included in the summation, and extends naturally to the full (non-jackknife) integrals with the omission of the $\left(1-\sum_Bw_{aB}w_{bB}\right)^{-1}$ prefactor, the disconnected term and $\omega^{ab}$ tensors. The $RR_a$ term may be estimated stochastically as
\beq\label{eq: RR_stochastic}
    \widehat{RR}_{a} &=& \frac{1}{N_\mathrm{pairs}}\left(\frac{N_g}{N_r}\right)^2\sum_{(i,j)\sim{}g^{(2)}}\left[N_r\frac{m_j}{p_j}\right]w^r_iw^r_j\Theta_a^{ij}.
\eeq
Since the $\omega_{ijkl}^{ab}$ jackknife weights depend on $RR_{aA}$ (through $w_{aA}$) these must be computed separately before we estimate the full integrals. As this only depends on pairs of points, little additional computational expense is required, making it feasible to compute this by counting all $N_r^2$ possible pairs without importance sampling (see Sec.\,\ref{subsec: pair_counting}). This gives the functional form
\beq\label{eq: RR_aAsum_def}
\widehat{RR}_{aA} = \left(\frac{N_g}{N_r}\right)^2\sum_{i\neq j}w^r_iw^r_j\Theta_a^{ij}q_{ij}^A
\eeq
where $i$, $j$ run over all random particles. The $RR_a$ term can simply be found via $\sum_ARR_{aA}$ in the unrestricted jackknife formalism (and can be compared to the above stochastic result as a useful test).

\subsubsection{Sampling Cell Selection Probabilities}\label{subsec: cell_probs}
The choice of the probabilities $p_i$ (giving the likelihood of selecting sampling cell $c_j$ containing particle $j$ from initial sampling cell $c_i$) allow us to optimize the performance of the \resub{Monte Carlo} estimators. We choose the probability for selecting a secondary cell at separation $\vec n$ from a primary cell to be proportional to 
\beq\label{eq: probIntegralInitial}
A(\vec{n})=\intop\intop W_{\vec{0}}\left(\vec{y}\right)W_{\vec{n}}\left(\vec{x}\right)K\left(\vec{x}-\vec{y}\right)d^{3}\vec{x}d^{3}\vec{y}\;,
\eeq
where $W_\vec{p}\left(\vec{y}\right)$ is the value, at spatial position $\vec y$, of a cubic sampling grid cell window function centered on position $\vec{p}$ of width $a$. Here $K(\vec r)$ is the kernel function, which is here integrated over both sampling grid cells.

Although the sampling cells are cubic, we may obtain much more tractable solutions to Eq.\,\ref{eq: probIntegralInitial} by treating them as spherical (keeping the volume fixed). In addition, by making the approximation that the window functions are spherical Gaussians rather than top-hat functions, the integral can be transformed to 
\beq\label{eq: probIntegralForXi}
    A_G(n)=\begin{cases}             \frac{1}{Rn\sqrt{\pi}}\exp\left(-\frac{n^2}{4R^2}\right)\int_0^\infty m K(m) \exp\left(-\frac{m^2}{4R^2}\right)\sinh\left(\frac{mn}{2R^2}\right)dm & n>0,\\
    \frac{1}{2\sqrt{\pi}R^3}\int_0^\infty m^2 K(m)\exp\left(-\frac{m^2}{4R^2}\right)dm & n=0,
    \end{cases}
\eeq
also assuming $K(\vec r)$ to be a function of $r=|\vec r|$ only. Note that importance sampling just requires the sampling distribution to be known, not to be perfect, thus these approximations do not compromise on accuracy (and lead to only a tiny reduction in efficiency). This is shown in appendix \ref{appen: probability_integrals} and can be computed numerically for arbitrary kernel functions.

In this paper, we will use two forms for the kernel function; $K(\vec r) = \xi(|\vec r|)$ and $K(\vec r) = |\vec r|^{-2}$, used for efficient importance sampling and uniform filling of all covariance matrix bins respectively (see Sec.\,\ref{subsec: algorithm_structure}). With the latter kernel, we have the semi-analytic result (derived in appendix \ref{appen: probability_integrals})
\beq\label{eq: probIntegralForR2}
    A_\mathrm{G}(n) = \begin{cases}
        \frac{2}{an}F\left(\frac{n}{a}\right) & n>0\\
        \frac{2}{a^2} & n=0
    \end{cases}
\eeq
for Dawson-F function $F$ \citep{DawsonF} (which can also be written in terms of confluent hypergeometric functions). This tends to $1/n^2$ at large radius but avoids infinities at small $n$. It is pertinent to note that simplifications in the form of the kernel and window function do not bias the computed covariance matrices, but just change the sampling strategy slightly. It is thus desirable to have a simple expression for $A(\vec{n})$. These forms set the probabilities $p_j$ used in the importance sampling estimator summations (e.g. Eqs. \ref{eq: normalized_jackknife_estimators}), which are computed before any Monte Carlo integration is performed.

In order to use the $\xi(r)$ kernel, we require that the input (radial) 2PCF is strictly positive over the full binning range, else some sampling cells will be erroneously excluded from the analysis. To ensure this we use the modified kernel
\beq\label{eq: xi_tilde_kernel}
\bar{\xi}(r) = \begin{cases}\xi(r) & \text{if }\xi(r)>10^{-2}\\ \frac{10}{r^{2.1}}& \text{else.}\end{cases}
\eeq
which was found to give efficient sampling of all sampling cells in the required range.

\subsubsection{Disconnected Term Evaluation}\label{subsec: disconnected_evaluation}
The disconnected part of $\hat C_{4,ab}^J$ (cf.\,Sec.\,\ref{subsec: disconnected_term}) is estimated in a different manner to the rest of the 4-point integral. For an arbitrary quad of points $(i,j,k,l)$, the $i-j$ and $k-l$ separations are constrained by the choice of bin (via the $\Theta_a^{ij}$ and $\Theta_b^{kl}$ functions), but the $i-k$ and $j-l$ separations are \textit{ab initio} unconstrained. If these are large, we will obtain a vanishingly small contribution to the connected 4-point terms since $\xi(\vec r)\rightarrow0$ for large $|\vec r|$. However $\xi_{ij}\xi_{kl}$ is independent of $|\vec r_i-\vec r_k|$ and $|\vec r_j-\vec r_l|$, making its evaluation more complex. 

As shown in appendix \ref{appen: disconnected}, the $\hat C_{x,ab}^J$ term should be small, yet the form of $\omega_{ijkl}^{ab}$ implies that this cancellation occurs via the balance of large positive contributions at small $i-k$ and $j-l$ separations with many small negative contributions on large scales (when $(i,j)$ and $(k,l)$ belong to different jackknife regions). Computationally, this only works if we consider the full range of $i-k$ and $j-l$ separations, yet we usually impose a cut-off at some large $i-k$ separation ($\sim 400$\,Mpc/h) for efficiency.

To ameliorate this, we use the alternative form of $C_{x,ab}^J$ (Eq.\,\ref{eq: disconnected_term}), which is computed from 2-point quantities ($EE_{aA}$), which are generally known to higher precision than the 4-point terms. Due to the finite sampling strategies used in computation of the $EE_{aA}$ terms, we still expect Poissonian fluctuations in the estimated values of $EE_{aA}-w_{aA}EE_a^J$, which reduce in amplitude as we evaluate the function at more points. For off-diagonal terms, this will lead to the estimate $\hat C_{x,ab}^J$ fluctuating about zero, yet for the leading diagonal (where $a = b$), we will not get cancellation, since the $\hat C_{x,aa}^J$ term involves a sum over $\left(EE_{aA}-w_{aA}EE_a\right)^2$ which is strictly positive. This can be removed by using two disjoint sets of survey points to compute $w_{aA}$ and $EE_{aA}$. This modifies the disconnected to the form
\beq
    \hat C_{x,ab}^J = \frac{1}{RR_a^JRR_b^J-\sum_BRR_{aB}RR_{bB}}\sum_A\left[EE_{aA}^{(1)}-w^{(1)}_{aA}EE_a^{J(1)}\right]\left[EE_{bA}^{(2)}-w_{bA}^{(2)}EE_b^{J(2)}\right]
\eeq
Here the labels (1) and (2) refer to estimates derived from different sets of random particles. The Poissonian fluctuations between the two $EE_{aA}-w_{aA}EE_{a}^J$ estimates are thus uncorrelated, removing the spurious diagonal term on average. We note that we must also compute $w_{aA}=RR_{aA}/RR_a^J$ for each set of particles individually to avoid obtaining substantial negative correlations between the two factors (as a large $EE_{aA}$ in one set of random particles would otherwise encourage a small $EE_{aA}$ in the other if $RR_{aA}$ were computed from all particle pairs including both random subsets). The $w_{aA}^{(n)}$ weights (for $n\in\{1,2\}$) are computed stochastically for the disconnected term, along with $EE_{aA}^{(n)}$. This is unlike that used for the rest of the analysis, but is appropriate since the 2-point terms are generally well-known and we expect the overall disconnected term to be small.

In practice, we assign each random particle to either subset 1 or 2 on initialization and computing contributions to the relevant $EE_{aA}$ count if both particles lie in the relevant bin and discarding else. This gives the estimator form for $EE_{aA}^{(n)}$ and $RR_{aA}^{(n)}$:
\beq\label{eq: disconnected_estimators}
    \widehat{EE}_{aA}^{(n)} &=& \frac{1}{N^{(n)}_\mathrm{pairs}}\left(\frac{N_g}{N_r}\right)^2\sum_{(i,j)\sim{}g^{(2,n)}}\left[N_r^{(n)}\frac{m_j^{(n)}}{p_j}\right]w_i^rw_j^r\Theta_a^{ij}\xi_{ij}\\\nonumber
    \widehat{RR}_{aA}^{(n)} &=& \frac{1}{N^{(n)}_\mathrm{pairs}}\left(\frac{N_g}{N_r}\right)^2\sum_{(i,j)\sim{}g^{(2,n)}}\left[N_r^{(n)}\frac{m_j^{(n)}}{p_j}\right]w_i^rw_j^r\Theta_a^{ij}
\eeq
where the superscripts $n$ refer to the relevant values for the $n$-th set of random particles. In addition, it is a fair assumption to assume $N_\mathrm{pairs}^{(1)}=N_\mathrm{pairs}^{(2)}$ as $N_\mathrm{pairs}/4$. 
In all tests, this term has been found to be small (several orders of magnitude smaller than the dominant $C^J_{ab}$ terms) but may be important for some particular choices of geometry and/or jackknife regions.

\subsubsection{Monte Carlo Estimator Scalings}\label{subsec: estimator-scalings}
It is pertinent to consider the dominant scalings of the above importance sampling integral estimators. These are as follows:
\begin{enumerate}
    \item \textbf{Number of random points, $N_r$}: Rescaling $N_r\rightarrow\beta N_r$, we expect the number of particles in each sampling cell to increase by a factor $\beta$ on average, thus $\av{m_j}\rightarrow\beta\av{m_j}$. From Eqs.\,\ref{eq: normalized_jackknife_estimators}\,\&\,\ref{eq: RR_stochastic}, it is clear that $\widehat RR_a$ and the covariance matrix estimators ($\hat C_{d,ab}$ and $\hat C^J_{d,ab}$) are invariant to this rescaling, which is as expected, since the random points are solely a computational aide.
    \item \textbf{Number of galaxies, $N_g$}: Rescaling $N_g\rightarrow \beta N_g$ has no effect on random particle terms, sampling cell probabilities or $\xi_{ij}$ (which is computed via normalized pair counts). Thus $\widehat RR_a\rightarrow\beta^2\widehat RR_a$ and $\hat C_{d,ab}\rightarrow\beta^{d-2}\hat C_{d,ab}$ (and similarly for the jackknife Monte Carlo estimators). Changing the number of galaxies in the survey volume thus reweights the relative contribution of the 2-, 3- and 4-point integral terms.
    \item \textbf{Average particle weight $\av{w_i}$}: Rescaling $w_i\rightarrow\beta w_i$ for all particles implies $w^r_i\rightarrow\beta w^r_i$ since we expect the random particle weight distributions to match that of the galaxies. Since all covariance matrix estimators involve four factors of $w^r$ in the numerator, and $\widehat RR_a$ involves two factors of $w^r$, we expect any dependence of the estimator on $\av{w_i}$ to vanish.
    \item \textbf{Number of jackknife regions, $N_J$}: Since $RR_a^J = RR_a$ (for the unrestricted jackknife) is independent of $N_J$, we must have $\widehat RR_{aA}\sim{N_J}^{-1}$ (as there are $N_J$ $RR_{aA}$ terms in total), and thus $w_{aA}\sim{N_J}^{-1}$. Inspection of the expansion of $\omega_{ijkl}^{ab}$ (Eq.\,\ref{eq: omega_tensor_expansion}), shows that this also scales as ${N_J}^{-1}$. In the jackknife integrals, the prefactor $\left(1-\sum_Bw_{aB}w_{bB}\right)^{-1}$ scales approximately as $\left(1-1/N_J\right)^{-1}$, thus $\hat C_{d,ab}^J\sim\left(N_J-1\right)^{-1}$ at leading order (which is exact if all jackknife regions are identical).
\end{enumerate}

\subsection{Shot-Noise Rescaling \& Precision Matrices}\label{subsec: shot_noise_rescaling}
Given the \resub{Monte Carlo} covariance matrix estimators, it remains to compute the optimal shot-noise rescaling parameter $\alpha$, which approximates non-Gaussianity in the model. Following \citet{2016MNRAS.462.2681O}, we do this by maximizing a likelihood based on the Kullback-Leibler (KL) divergence \citep{kullback1951} between our estimate of the jackknife covariance matrix, $\hat C^J(\alpha)$, and the data-derived estimate, $\hat C^J_D$, which uses the 2PCF estimates from only a single dataset, without reference to mocks. Denoting the theoretical precision matrix as $\hat \Psi^J(\alpha)$ (equal to the inverted $\hat C^J(\alpha)$ matrix in the noiseless case), we formulate the likelihood $\mathcal{L}_1$ as 
\beq\label{eq: KL_likelihood}
    -\log\mathcal{L}_1(\alpha) = 2 D_{KL}\left(\hat\Psi^J(\alpha),\hat C_D^J\right) = \text{trace}\left[\hat\Psi^J(\alpha)\hat C^J_D\right] - \log\det \hat C^J_D - \log\det \hat\Psi^J(\alpha) - n_\mathrm{bins}
\eeq
for a total of $n_\mathrm{bins}$ bins.\footnote{We use the $\mathcal{L}_1$ likelihood here rather than $\mathcal{L}_2$ (with $-\log\mathcal{L}_2 = 2 D_{KL}\left(\hat\Psi^J_D,\hat C^J(\alpha)\right)$, since the latter requires inversion of the singular jackknife data covariance matrix.} It is pertinent to note that this likelihood does not account for noise in the fitting matrices, which can cause a small bias, as will be discussed in future work. Here, $\alpha$ is determined by maximizing $\mathcal{L}_1$ numerically, which requires 
\beq\label{eq: KL_likelihood_max}
    &{}& \mathrm{trace}\left[\left(\hat C_D^J-\hat C^J(\alpha)\right)\frac{\partial \hat\Psi^J(\alpha)}{\partial \alpha}\right] = 0\\\nonumber
    &\Rightarrow& \mathrm{trace}\left[\left(\hat C_D^J-\hat C^J(\alpha)\right)\hat\Psi^J(\alpha)(\hat C_3^J+2\alpha \hat C_2^J)\hat\Psi^J(\alpha)\right] = 0
\eeq
utilizing the identities
\beq
    \frac{\partial}{\partial{\alpha}}\log\det \hat\Psi^J(\alpha) &=& \mathrm{trace}\left(\hat\Psi^J(\alpha)^{-1}\frac{\partial}{\partial\alpha}\hat\Psi^J(\alpha)\right)\\\nonumber
    \frac{\partial \hat\Psi^J(\alpha)}{\partial \alpha} &=& -\hat\Psi^J(\alpha)\frac{\partial \hat C^J(\alpha)}{\partial \alpha}\hat\Psi^J(\alpha). 
\eeq
As shown in \citet{2019MNRAS.487.2701O}, a simple inversion of a noisy covariance matrix yields a biased estimate of the precision matrix $\Psi$. For a general (covariance) matrix $C$, a bias-corrected estimator of $\Psi = C^{-1}$is given by 
\beq\label{eq: bias_corrected_precision}
    \hat \Psi &=& \left(\mathbb{I}-\tilde D\right) \hat C^{-1} \\\nonumber
    \tilde D &=& \frac{n_\mathrm{samples}-1}{n_\mathrm{samples}}\left[-\mathbb{I}+\frac{1}{n_\mathrm{samples}}\sum_m\hat C_{[m]}^{-1}\hat C_m\right]
\eeq
computed using $n_\mathrm{samples}$ independent estimates of the matrices, denoted $\hat C_{m}$, with $\hat C_{[m]}$ indicating the mean of the other $n_\mathrm{samples}-1$ samples, excluding $\hat C_m$. This requires multiple model covariance matrix estimates, which are easy to obtain using our stochastic sampling algorithm (Sec.\,\ref{sec: algorithm_design}. This is used to compute the full and jackknife precision matrices here, free from quadratic bias.

For a matrix $M$ following Wishart statistics, computed from $n_\mathrm{samples}$ samples in $n_\mathrm{bins}$, we have the general result (\citealt{wishart28}; see also \citealt{2007A&A...464..399H} for application to cosmology);
\beq
    M^{-1} = \left(1-D\right)\av{\hat M^{-1}}, \qquad
    D = \frac{n_\mathrm{bins}+1}{n_\mathrm{samples}-1}.
\eeq
Comparison with Eq.\,\ref{eq: bias_corrected_precision} allows us to define the \textit{effective number of mocks}, $n_\mathrm{eff}$ for a general matrix, which would be equal to $n_\mathrm{samples}$ in the case of Wishart noise. Using the mean determinant per mode $\left|\tilde D\right|^{1/n_\mathrm{bins}}$ as a proxy for $D$;
\beq\label{eq: n_eff}
    n_\mathrm{eff} = \frac{n_\mathrm{bins}+1}{\left|\tilde D\right|^{1/n_\mathrm{bins}}}+1.
\eeq
This differs from the definition of \citet{2016MNRAS.462.2681O,2019MNRAS.487.2701O}, which uses the variance of $\hat M$ for off-diagonal elements, and is chosen to avoid the former ambiguities in the choice of bin. Here this is performed for $\hat C_{ab}$ and $\hat C_{ab}^J$ in post-processing to assess the precision of the computed matrices.

\subsection{Correlation Function Estimation}\label{subsec: corr_functions}
A key input to the covariance matrix integrals is the estimated 2PCF $\xi(r,\mu)$ for $r = |\vec r|$ and angular coordinate $\mu = \cos\theta$, where $\theta$ is the angle of a pair of galaxies from (a) their mean line of sight position (for a non-periodic dataset) or (b) the $z$ axis (for a periodic simulation). In addition, we require estimates of the 2PCF in each jackknife region, in order to compute the data jackknife covariance matrix which can be compared to theory to estimate the shot-noise rescaling $\alpha$. To compute $\xi$ from a data-set, we use the \citet{1993ApJ...412...64L} estimator, which defines a full-survey 2PCF in bin $a$ as
\beq\label{eq: LS_correlation}
    \hat{\xi}_a^{LS} = \frac{\widetilde{DD}_a - 2\widetilde{DR}_a + \widetilde{RR}_a}{\widetilde{RR}_a}
\eeq
where $\widetilde{DD}_a$, $\widetilde{DR}_a$ and $\widetilde{RR}_a$ represent normalized auto- and cross-pair counts for the galaxy $D$ and random particle $R$ fields. These are defined as
\beq\label{eq: weighted_pair_count}
    \widetilde{FG}_a = \frac{\sum_{i\in F}\sum_{j\in G}\Theta^{ij}_aw^F_iw^G_j}{\left(\sum_{i\in F}w_i^F\right)\left(\sum_{j\in G}w_j^G\right)}
\eeq
for $F,G\in\{D,R\}$, weighting by the product weight $w^F_iw^G_j$ and summing over all particles in each field. The normalization accounts for differences in number of particles in each dataset. 
For the unrestricted jackknife correlations, the estimator is modified to 
\beq\label{eq: jackknife_correlation_LS}
    \hat{\xi}_{aA}^{LS,J} &=& \frac{\widetilde{DD}_{aA} - 2\widetilde{DR}_{aA} + \widetilde{RR}_{aA}}{\widetilde{RR}_{aA}}
\eeq
where the jackknife pair counts are defined as
\beq\label{eq: weighted_pair_count_jk}
    \widetilde{FG}_{aA} = \frac{\sum_{i\in F}\sum_{j\in G}\Theta^{ij}_aw_i^Fw_j^Gq_{ij}^A}
    {\left(\sum_{i\in F}w_i^F\right)\left(\sum_{j\in G}w_j^G\right)}.
\eeq
These estimates are then used to compute the data jackknife covariance matrix via Eq.\,\ref{eq: C_jackknife_estimator}. Notably, we normalize the jackknife pair counts by the same factors as for the full-survey pair counts, \textit{not} the summed weights of the relevant jackknife regions. This means that our $\hat\xi_{aA}^{LS,J}$ estimates are not truly representative of the 2PCF in the unrestricted jackknife $A$, with discrepancies arising since we ignore differences in the ratio of galaxies to random particles between different jackknives (which occur both due to the small number statistics and large scale correlations). Here, this form is preferred since it can be simply compared against our theoretical jackknife estimate, $\hat C^J$, which assumed a uniform galaxy-to-random ratio for each jackknife. 

Using the Landy-Szalay 2PCF estimators above, we obtain $\hat\xi$ averaged across radial and angular bins. For out covariance matrix estimators (Eqs.\ref{eq: C234FullIntegrals}\,\&\,\ref{eq: NewC234Definitions}), we instead require knowledge of $\hat\xi(r,\mu)$ at arbitrary radii and angles, thus we must convert from a binned to a continuous function. A simple solution would be to use the bin-averaged estimates $\hat\xi_a$ as the values of $\hat\xi(r,\mu)$ at the bin-centers (denoted $r_{a_1}$ and $\mu_{a_2}$) and use linear interpolation to convert this into continuous space. However, this introduces a non-negligible bias, since the true values of $\hat\xi(r_{a_1},\mu_{a_2})$ are not equal to $\hat\xi_a$. In this paper, as in \citet[Sec.\,3.2][]{2016MNRAS.462.2681O}, we adopt a different approach, computing a corrected set of 2PCF values, $\hat\xi_\mathrm{corr}(r_{a_1},\mu_{a_2})$. A continuous function $\hat\xi(r,\mu)$ is computed from these via linear interpolation which, by construction, reproduces the binned $\hat\xi_a$ values when averaged over the original bins.

To compute $\hat\xi_\mathrm{corr}$, we adopt an iterative procedure, gradually refining the estimates of the 2PCF values at the bin centers. Given a continuous function from estimates, $\xi^{(m)}_\mathrm{corr}(r_{a_1},\mu_{a_2})$, the bin-averaged 2PCF is given by
\beq
    \xi^{(m)}_{a,\mathrm{binned}} = \frac{1}{RR_a}\int d^3\vec r_i d^3\vec r_j\,n(\vec r_i)n(\vec r_j)w(\vec r_i)w(\vec r_j)\Theta_a(\vec r_i-\vec r_j)\xi^{(m)}_\mathrm{corr}(\vec r_i-\vec r_j).
\eeq
These are computed in the same manner as the $C_{2,ab}$ integrals (described below), with a very small level of noise. This is then compared to the true value $\xi^\mathrm{true}_a$ (as found via the Landy-Szalay estimator) which is used to find the next estimate as
\begin{eqnarray}\label{eq: refined_xi}
    \xi^{(m+1)}_\mathrm{corr}(r_{a_1},\mu_{a_2}) &=& \frac{\xi^\mathrm{true}_a}{\xi^{(m)}_{a,\mathrm{binned}}}\xi^{(m)}_\mathrm{corr}(r_{a_1},\mu_{a_2}).
\end{eqnarray}
Starting from an initial estimate of $\xi^{(0)}_\mathrm{corr}(r_{a_1},\mu_{a_2}) = \xi_a^\mathrm{true}$, we are able to obtain sub-percent agreement between $\xi^{(m)}_{a,\mathrm{binned}}$ and $\xi_a^\mathrm{true}$ after $\sim10$ iterations.

\section{Algorithm Overview}\label{sec: algorithm_design}
We here review the structure of the \texttt{RascalC} code used to \resub{implement the Monte Carlo estimators} and perform all necessary pre- and post-processing.\footnote{\url{https://github.com/oliverphilcox/RascalC}} 

\subsection{Pair Counting}\label{subsec: pair_counting}
To evaluate the jackknife covariance integrals (Eqs.\,\ref{eq: NewC234Definitions} or \ref{eq: C234JackIntegrals}), we require an estimate of the $w_{aA}$ weights to allow computation of the $\omega_{ijkl}^{ab}$ tensor. This in turn requires knowledge of the $RR_{aA}$ (and hence $RR_a^J=RR_a$ functions), which are 6-dimensional integrals, depending on a pair of points in 3-dimensional space. In addition, the $RR_a$ functions are an important normalization factor for both full and jackknife covariance matrices. Although these could be computed stochastically (as in Eq.\,\ref{eq: RR_stochastic}), it is relatively simple to compute them exhaustively, as a weighted sum over all pairs in each bin obeying the correct unrestricted jackknife criterion. 

In practice, these are computed using the \texttt{corrfunc} code\footnote{\url{https://corrfunc.readthedocs.io/en/master/}} \citep{2017ascl.soft03003S}, which efficiently counts all possible pairs of particles in two input fields, given a set of input $(r,\mu)$ bins. As in the \resub{importance sampling} estimators for the 2-, 3- and 4-point matrices (Eqs. \ref{eq: normalized_jackknife_estimators}), we must multiply the \texttt{corrfunc} $RR_a$ and $RR_{aA}$ estimates by a factor $(N_g/N_r)^2$ to account for the different numbers of galaxies and random points (as in Eq. \ref{eq: RR_aAsum_def}). 

For the unrestricted jackknife counts in a jackknife region $A$, the jackknife-pair counts $RR_{aA}$ is given by a pair count using the cross-correlation of the pairs in the full survey with those in the jackknife region $A$, weighted by the product of particle weights $w_iw_j$. This has the desired effect of weighting assigning a weight of unity to pairs entirely in the jackknife and half for those with only one-half in the jackknife. This can be done in $N_J$ iterations of \texttt{corrfunc}, and does not require a significant increase in computation time compared to the $RR_a$ computation. In addition, we may simply co-add the $RR_{aA}$ estimates to find $RR_a^J=RR_a$ for each bin in this jackknife formalism.

In addition, \texttt{corrfunc} pair counts may be used to define the input full-survey and jackknife 2PCFs (but can be specified separately, if required). This simply uses the Landy-Szalay formalism (Eqs.\,\ref{eq: LS_correlation}\,\&\,\ref{eq: jackknife_correlation_LS}), requiring pair counts between data and random-particle fields $D$ and $R$. For the $\hat\xi_a$ estimation, this is done as for $RR_a$, with normalization now given as the summed weights in the two input fields. For jackknife pair counts $FG_{aA}$ of two fields $F,G\in\{D,R\}$, we use the mean pair counts from the entirety of field $F$ and the jackknife region $A$ of field $G$ and from the entirety of field $G$ and the jackknife region $A$ of survey $F$, if $F\neq G$ (i.e. for $DR$ pair counts). We note that the binning used for the $\hat\xi_{aA}$ functions should match that of the covariance matrices, but may differ from the input $\hat\xi_a$. 

\subsection{\texttt{RascalC} Algorithm Structure}\label{subsec: algorithm_structure}
As described in Sec.\,\ref{sec: matrix_evaluation}, we compute the covariance matrix estimates by summing over sets of four particles (`quads'). The basic structure may be summarized as follows.

First, the survey region is discretized into a number of cubic sampling cells, to allow an efficient implementation of the Monte Carlo importance sampling (as described in Sec.\,\ref{subsec: grids}), and random particles are read-in, assigning the $i$-th particle to sampling cell $c_i$ (which can contain than one particle). For later estimation of the disconnected $\hat C_{x,ab}^J$ term, each particle is assigned to random subclass 1 or 2 (Sec.\,\ref{subsec: disconnected_evaluation}). 

Given the sampling cell size, the transition probabilities $p_j$ between cell $c_j$ and $c_i$ (Sec.\,\ref{subsec: cell_probs}) are computed both using the $\bar\xi(r)$ kernel (Eq.\,\ref{eq: probIntegralForXi}) and the $1/r^2$ kernel (Eq.\,\ref{eq: probIntegralForR2}). Since these depend only on inter-cell separations, they may be precomputed without knowledge of the individual particle positions. Here, numerical integration is performed for the $\bar\xi(r)$ kernel via the \texttt{cubature} C\texttt{++} package.\footnote{\url{https://github.com/stevengj/cubature}}, up to a maximum separation corresponding to the cut-off beyond which correlations are set to zero. When using a large enough value for this cut-off, the truncation was found to give no significant impact on the overall result. 

At this point, the 2PCF $\xi(r,\mu)$ is computed from the input binned function $\hat\xi_a$ as described in \,Sec.\,\ref{subsec: corr_functions}, and the bin-averaging integration is performed stochastically, by drawing pairs in the same manner as with the 2-point terms below. This refinement cannot be performed alongside the 2-point integral computation, since $\xi(r,\mu)$ is an important part of the covariance matrix integrands.

Following such pre-processing, a stochastic computation of the 2-, 3- and 4-point integrals is performed via the Monte Carlo estimators of Eqs.\,\ref{eq: normalized_jackknife_estimators}. This is done over a number of independent epochs $N_\mathrm{epochs}$, each of which utilizes each random particle in turn as the primary particle, and may be run on separate cores. By splitting up the computation into shorter epochs, we produce independent estimates of the integrals, allowing better estimation of the precision matrix (cf.\,Sec.\,\ref{subsec: shot_noise_rescaling}) and the effective number of samples ($n_\mathrm{eff}$) in the code to be assessed. In each epoch, we adopt the following procedure (chosen to ensure minimal re-computation of quantities such as the interpolated $\xi(r,\mu)$):
\begin{enumerate}
    \item Iterate over each non-empty cell (denoted $c_i$) in the sampling grid. This ensures that every random particle (in a total of $N_\mathrm{cells}$ cells) is used in each epoch.
    \item Randomly select a second sampling cell ($c_j$) from $c_i$ with probability $p_j$, using the pre-computed probability grid. This is performed using the Walker-Vose alias method (\citealt{walker1974new,walker1977efficient,vose1991linear}, in the implementation of Joachim Wuttke\footnote{\url{apps.jcns.fz-juelich.de/man/ransampl.html}}) and selects one realization from a discrete set of possibilities with pre-assigned probabilities, here the grid of neighbouring sampling cells. To draw the $c_j$ cell, we use the $1/r^2$ kernel (with $p_j$ defined by Eq.\,\ref{eq: probIntegralForR2}), to ensure that all bins are filled roughly uniformly, truncating at the maximum radial bin size. If the selected cell lies in the survey region and is non-empty, we pick a single particle ($j$) from the $c_j$ cell and iterate over all $i$ particles in the $c_i$ cell. For each $(i,j)$-pair of particles, we do the following;
    \begin{itemize}
        \item Compute the correlation bin $a$ from the $(r,\mu)$ separation of the particles.
        \item Evaluate the correlation function at the particle separation ($\xi_{ij}$) via linear interpolation of the input 2PCF.
        \item Find the contributions to the 2-point integrals $RR_a$, $\hat C_{2,ab}$ and $\hat C_{2,ab}^J$ and add to the sum in the relevant bin.  (The stochastic estimation of $RR_a$ is not used directly, but it may be compared with the value obtained from \texttt{corrfunc} to ensure that the importance sampling is working as expected.) We do not normalize by $RR_a$ or $(1-\sum_Bw_{aB}w_{bB})$ yet.
        \item If the particles are in the same random subclass $n$, add the contributions to the $EE^{(n)}_{aA}$ and $RR^{(n)}_{aA}$ integrals (via Eq.\,\ref{eq: disconnected_estimators}).
    \end{itemize}
    This step is repeated $N_2$ times for each primary sampling cell $c_i$. 
    \item For each $c_j$ cell, pick a third sampling cell $c_k$ from $c_i$ with probability $p_k$, now selecting via the $\bar\xi(r)$ kernel (Eq.\,\ref{eq: probIntegralForXi}) for efficient importance sampling over all $i-j$ and $j-k$ bins. If a valid sampling cell $c_k$ is picked, a random particle ($k$) is chosen from it and the contributions to the 3-point integrals $\hat C_{3,ab}$ and $\hat C_{3,ab}^J$ are computed for the chosen $(i,j,k)$-triples of particles (iterating over all $i$ particles in cell $c_i$, but a single $j$ and $k$ particle) in the relevant $i-j$ and $j-k$ bins $a$ and $b$. This is repeated $N_3$ times per $c_j$ cell.
    \item For each $c_k$, choose a fourth sampling cell $c_l$ from $c_j$ via a $\bar\xi(r)$ with probability $p_l$ giving a total probability $p_jp_kp_l\sim\xi_{ik}\xi_{jl}$, matching that of the 4-point integrand. If the sampling cell is non-empty, we pick a random particle representative ($l$) and compute 4-point terms $\hat C_{4,ab}$ and $\hat C_{4,ab}^J$ for each $(i,j,k,l)$-quad (again using all $i$ particles in $c_i$ but a single particle from each of the $c_j$, $c_k$ and $c_l$ cells). This is repeated $N_4$ times per cell, and each iteration only involves a single $\xi(r,\mu)$ interpolation, for $\xi_{jl}$ (since the $i$, $j$ and $k$ particles are held constant for this loop). 
\end{enumerate}

The integers $N_2$, $N_3$ and $N_4$ can be varied to allow for manageable computation times and for the precision of each matrix to be tuned individually. In each epoch, we attempt to draw $N_\mathrm{cells}N_2N_3N_4$ quads of sampling cells, with up to $N_\mathrm{r}N_2N_3N_4$ quads of particles utilized. (Note that this is an upper bound since many quads are discarded due to empty cells and particle pairs outside the binning ranges). In our C\text{++} implementation, when applied to the BOSS DR12 survey (Sec.\,\ref{subsec: qpm_cov_single}), quads of sampling cells are selected at $\sim10^{7}$ quads/second/core, and quads of particles accepted at $\sim5\times10^{6}$ quads/second/core, though we note this is survey geometry dependent.

At the end of each epoch, the summations are added to the current global estimates of the covariance matrices and the disconnected terms evaluated from the $EE_{aA}$ and $RR_{aA}$ estimates. Normalization is performed by dividing by the number of pairs/triples/quads of particles which the algorithm attempts to use, e.g. by $N_\mathrm{quads} = N_\mathrm{epochs}N_rN_2N_3N_4$ for quads.\footnote{Note that to ensure correct use of the generated sampling cell probability grids we must normalize by the number of \textit{attempted} particles (including those rejected for being out of the survey region and in invalid bins), rather than the number actually accepted by the code.} We further normalize by the pair-counts $RR_a$ and jackknife weight normalization $(1-\sum_Bw_{aB}w_{bB})$ to approximate the covariance integrals (Eqs. \ref{eq: C234FullIntegrals} \& \ref{eq: NewC234Definitions}).

Once all submatrices are estimated, the overall covariance matrices $\hat C_{ab}$ and $\hat C_{ab}^J$ are reconstructed in Python. Given estimates of the jackknife 2PCFs, $\hat\xi_{aA}^J$, the shot-noise rescaling parameter is computed, as in Sec.\,\ref{subsec: shot_noise_rescaling}, and we output the rescaled full and jackknife matrices, as well as their associated precision matrix forms and effective number of matrix samples (via Eqs.\,\ref{eq: bias_corrected_precision}\,\&\,\ref{eq: n_eff}).

\subsection{Measures of Convergence}\label{subsec: convergence}
We can dynamically estimate the gradual convergence of the matrix estimates with increasing $N_\mathrm{epoch}$ using the \textit{Frobenius norms} of the stochastic matrix estimators, following \cite{2010arXiv1010.3866C}. This is defined for $n\times{}n$ matrices $A$ and $B$ as 
\beq\label{eq: FrobeniusNorm}
    \mathrm{norm}_{F}(A,B) = \sqrt{\sum_{i=1}^n\sum_{j=1}^n(A_{ij}-B_{ij})^2}.
\eeq
Here we compare the $\hat C_d$ matrix estimates before (denoted $\hat C_{d,ab}$) and after (denoted $\hat C_{d,ab}'$) inclusion of additional epoch(s) of data, computing the fractional difference as 
 \beq
    \mathrm{diff}(\hat C_d,\hat C_d') = \frac{\mathrm{norm}_F(\hat C_d,\hat C_d')}{\mathrm{norm}_F(\hat C_d,\hat C_d)}.
\eeq
This allows us to see how the matrix estimates are converging and can be used to halt the algorithm when some desired precision is reached. We found the Frobenius norm to be more useful than the KL divergence for evaluating convergence during the algorithm's runtime, since the latter approach requires the matrix estimates to have no negative eigenvalues (to give positive matrix determinants) which is not guaranteed for small runtimes. 

After the algorithm is complete, a more concrete determination of the matrix convergence is provided via the effective number of mocks, computed from the variation of individual matrix estimates, as in Sec.\,\ref{subsec: shot_noise_rescaling}. In most common usages of covariances (e.g. in the determination of Fisher information matrices or model testing with $\chi^2$ minimization) we require them instead in precision matrix form. This requires a highly converged covariance matrix to reduce noise in the precision matrix. 

An additional measure of convergence is obtained from considering the eigenvalues of the computed matrices. To compute estimates of the precision matrices $\hat\Psi_{ab}$ and $\hat\Psi^J_{ab}$, we must invert our covariance matrices $\hat C_{ab}$ and $\hat C^J_{ab}$. Numerically, this requires that the eigenvalues of the (symmetric) total matrices be positive, but random noise in the matrices may oppose this. In particular, we expect the high-dimensional 4-point matrices to be the least well constrained, and we find that there exist negative eigenvalues in these matrices with small run-times. A condition for matrix inversion (for both full and jackknife matrices) is hence
\beq
    \operatorname{min\,eig}(\hat C_{4,ab})+\alpha\operatorname{min\,eig}(\hat C_{3,ab})+\alpha^2\operatorname{min\,eig}(\hat C_{2,ab}) \geq 0
\eeq
where $\operatorname{min\,eig}(A)$ represents the minimum eigenvalue of $A$. This uses Weyl's inequality
\beq
    \operatorname{min\,eig}(A)+\operatorname{min\,eig}(B)\geq \operatorname{min\,eig}(A+B)
\eeq
for Hermitian matrices $A,B$. Assuming the 2- and 3-point matrices to be well converged (with positive eigenvalues) and $\alpha\geq1$, a necessary condition for the matrix inverse to exist is hence
\beq
    \operatorname{min\,eig}(\hat C_{4,ab})\geq-\operatorname{min\,eig}(\hat C_{2,ab}).
\eeq
This provides an important convergence test in post-processing.

\section{Results}\label{sec: results}
Here we show the usage of our \texttt{RascalC} covariance matrix estimation code both via comparison with the \texttt{Rascal} Python code and through application to a suite of mock galaxies. This demonstrates the consistency of our code with previous approaches, as well as showing its utility in real settings. Throughout the section we will use linear binning for $r$ and $\mu$ in the covariance matrix with $\Delta r=4\,h^{-1}\,\text{Mpc}$, $\Delta\mu=0.1$ and $r\in[40,180]\,h^{-1}\,$Mpc, giving a total of 35 radial and 10 angular bins. \resub{Whilst this is a greater number of bins than used in most current analyses, it allows for simple comparison with previous works \citep{2016MNRAS.462.2681O,2019MNRAS.487.2701O} as well as to stringently test out analysis, since using more bins leads to greater off-diagonal matrix contributions and slower convergence.} The binning adopted for the input 2PCF $\xi(r,\mu)$ varies between tests.

\subsection{Comparison with \texttt{Rascal}}\label{subsec: rascal_comparison}
In order to demonstrate the validity of our new covariance matrix estimation approach, we should compare the results to those from the previous code, \texttt{Rascal}, which used a mask file and an estimate of the galaxy distribution $n(z)$ to sample the integrals, rather than using random particle files. Here, we compare covariance matrices from a \texttt{Rascal} run (described in \citealt{2019MNRAS.487.2701O}) to an associated \texttt{RascalC} run, computed using the same $N_g$ and input 2PCF $\xi$. This correlation function uses narrow bins of $\Delta r=1\,h^{-1}\,\text{Mpc}$, $\Delta \mu = 0.01$ for $r\in[0,180]\,h^{-1}\,$Mpc, with additional smoothing applied. The \texttt{RascalC} matrix was computed by attempting to sample $N_\mathrm{quads}=2\times10^{12}$ quads of particles (including those rejected e.g. by being outside the survey region), with a total integration time of $\sim60$ core-hours. Since the two codes used different jackknife formalisms, we only compare the full matrices here.

We estimate the level of noise in each run via the effective number of mocks, $n_\mathrm{eff}$ here computed via the bias in the off-diagonal precision matrix ($\Psi_{ab}$) elements;
\beq\label{eq: n_eff_old}
    n_\mathrm{eff} = n_\mathrm{bins}+\left[\operatorname{var}\left(\frac{\Psi_{ab}}{\sqrt{\Psi_{aa}\Psi_{bb}}}\right)\right]^{-1}
\eeq
for $r_a\geq142\,h^{-1}$\,Mpc and $r_b\leq82\,h^{-1}$\,Mpc as in \citet{2016MNRAS.462.2681O}.\footnote{We adopt this approach to measure $n_\mathrm{eff}$ here rather than the previously used subsampled $\tilde{D}$ matrix based technique (Eq.\,\ref{eq: n_eff}), since we do not have a large number of independent subsamples for the \texttt{Rascal} run. In general these estimates give broadly similar results.} We note that this will vary as a function of the shot-noise rescaling parameter, $\alpha$, due to the different weightings of matrix terms (with smaller $\alpha$ giving greater weight to the more noisy $C_4$ term). Here, the level of noise in the \texttt{RascalC} runs is far smaller than for \texttt{Rascal} (with $n_\mathrm{eff}\sim10^6$ and $n_\mathrm{eff}\sim10^4$ respectively), so is considered smooth here. 

One obstacle to determining whether the discrepancies between $C_\mathrm{Rascal}$ and $C_\mathrm{RascalC}$ are consistent with noise is that the noise on each matrix is not Wishart-distributed. Absent a detailed understanding of that noise, we can look for consistency at the order-of-magnitude level by modeling the noise as Wishart, in which case the expected KL divergence between the two matrices is
\beq\label{eq: KL_div_from_n_eff}
    D_{KL,\mathrm{expected}} = \frac{n_\mathrm{bins}(n_\mathrm{bins}+1)}{4n_\mathrm{eff}(C_\mathrm{Rascal})},
\eeq
as shown in appendix \ref{appen: KL-div}, where we use $n_\mathrm{eff}(C_\mathrm{Rascal})$ as the Wishart sample size, appropriate in the $n_\mathrm{eff}(C_\mathrm{Rascal})\ll n_\mathrm{eff}(C_\mathrm{RascalC})$ limit (as here). This may be compared to the measured KL divergence via the standard formula
\beq\label{eq: KL_divergence}
    D_{KL,\mathrm{measured}} = \frac{1}{2}\left[\mathrm{tr}(\Psi_\mathrm{RascalC}C_\mathrm{Rascal})-n_\mathrm{bins}-\operatorname{log\,det}(\Psi_\mathrm{RascalC})-\operatorname{log\,det}(C_\mathrm{Rascal})\right]
\eeq
here inverting the \texttt{RascalC} run due to its greater smoothness. If the expected KL divergence is far smaller than observed, we can posit that the differences are not consistent with noise on the \texttt{Rascal} matrix alone. 

\begin{figure}
\centering
\begin{minipage}[t]{.4\textwidth}
\vspace{0pt}
    \centering
\captionsetup{type=table}
\begin{tabular}{c|cc}
          $\alpha$ & $D_{KL,\mathrm{expected}}$ & $D_{KL,\mathrm{measured}}$\\
          \hline
            0.9 & 2.79 & 1.90 \\
            1.0 & 1.85 & 1.33 \\
            1.1 & 1.28 & 0.96 \\
            1.2 & 0.92 & 0.71 
    \end{tabular}
    \caption{Comparison of true KL divergences (Eq.\,\ref{eq: KL_divergence}) between \texttt{Rascal} and \texttt{RascalC} precision matrix estimates and their expected values given the matrix noise (Eq.\,\ref{eq: KL_div_from_n_eff}), as a function of shot-noise rescaling parameter $\alpha$. If matrix differences are due to noise alone, we expect order-of-magnitude consistency between the two sets of results, as seen here. This indicates no significant deviations between the codes at a precision level corresponding to $\sim 10^4$ mocks. As expected, the KL divergences decrease as $\alpha$ increases, which gives greater weight to more well converged terms. A comparison of the two precision matrices at $\alpha=1$ is shown in Fig.\,\ref{fig: RascalComparisonResidual}.}
    \label{tab: KL_differences}    
\end{minipage}%
\hfill
\begin{minipage}[t]{.5\textwidth}
\vspace{0pt}
  \centering
  \includegraphics[width=0.8\linewidth]{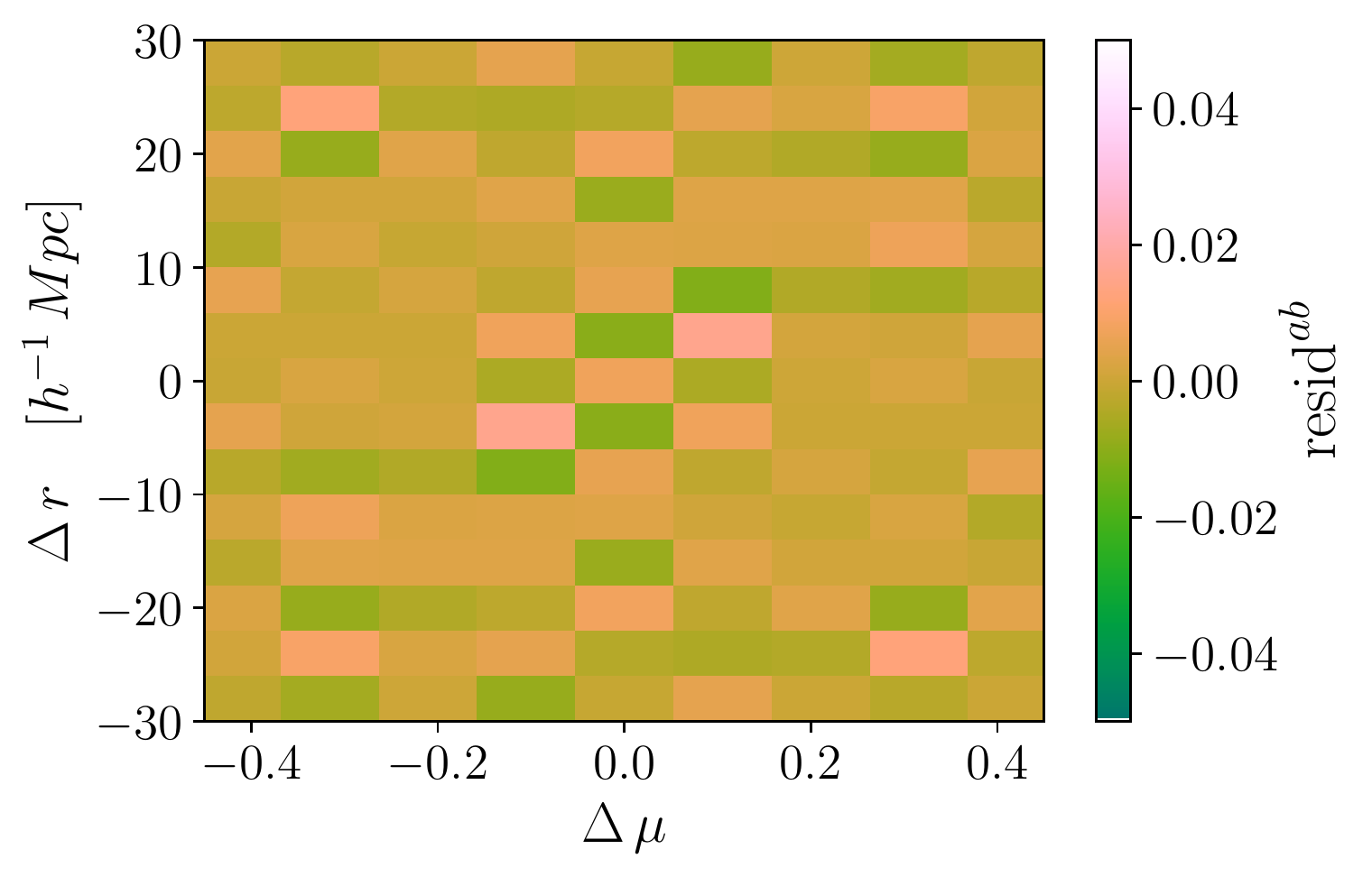}
  \caption{Stacked matrix residual between two whitened precision matrices computed by the original \texttt{Rascal} and new \texttt{RascalC} codes, using shot-noise rescaling $\alpha=1$. The residual is defined in Eq.\,\ref{eq: residual_matrix}, and each $(\Delta r, \Delta \mu$) cell indicates a mean of all residual matrix elements ($\mathrm{resid}^{ab}$) satisfying $r_a-r_b=\Delta r$ and $\mu_a-\mu_b=\Delta\mu$. This structure appears to be random and thus consistent with noise, indicating no obvious systemfatic differences between the two codes.}
  \label{fig: RascalComparisonResidual}
\end{minipage}
\end{figure}

The values of $D_{KL,\mathrm{expected}}$ and $D_{KL,\mathrm{measured}}$ are given in Tab.\,\ref{tab: KL_differences} for a range of values of $\alpha$, and show the expected decrease in $D_{KL}$ with increasing $\alpha$. In addition, we note order-of-magnitude consistency between the expected and measured KL divergences across the range of $\alpha$ tested. (Since the matrices are not Wishart distributed, different estimators of $n_\mathrm{eff}$ yield somewhat different results, thus we do not expect perfect consistency here). This similarity implies that the KL divergence between the two matrices can be attributed to noise on the \texttt{Rascal} matrix, giving no evidence for a systematic deviation between the codes (and underlying sampling algorithms) at the noise-level of the \texttt{Rascal} run ($n_\mathrm{eff}\sim10^4$).

To show this graphically, in Fig.\,\ref{fig: RascalComparisonResidual} we display the stacked residual matrix between \texttt{Rascal} and \texttt{RascalC} runs with $\alpha=1$, where we define the (whitened) residual matrix as 
\beq\label{eq: residual_matrix}
    \mathrm{resid}^{ab} = \frac{\Psi_\mathrm{Rascal}^{ab}-\Psi_\mathrm{RascalC}^{ab}}{r_ar_b}.
\eeq
Here we stack all matrix elements with a given $\Delta\mu=\mu_a-\mu_b$ and $\Delta r=r_a-r_b$ together to aid interpretation. There is no obvious structure to this matrix, hence we note no clear systematic differences between the two precision matrices, to this level of noise. We thus conclude that the two codes give comparable results at least up to $n_\mathrm{eff}\sim10^4$.

\subsection{Covariance Matrices of QPM Mocks}\label{subsec: qpm_cov_single}
\subsubsection{Single Mock Analysis}
To test our analysis we initially apply the jackknife covariance matrix formalism to a mock galaxy dataset, using a single Quick Particle Mesh (QPM) simulation \citep{2014MNRAS.437.2594W}, which emulates the NGC CMASS dataset \citep{2013AJ....145...10D} from Data Release 12 of the Baryon Oscillation Spectroscopic Survey \citep[BOSS;][]{2015ApJS..219...12A,2017MNRAS.470.2617A}, part of the Sloan Digital Sky Survey III \citep[SDSS-III;][]{2011AJ....142...72E}. This consists of a set of 642051 galaxy and 32292068 (with $N_\mathrm{r}=50N_g$) random particle positions, with respective FKP weights \citep{1994ApJ...426...23F} which define $w(\vec r)$. Positions are converted to a Cartesian frame assuming the cosmology $\{\Omega_m=0.29,\Omega_k=0,w_\Lambda=-1\}$ \citep{2018MNRAS.477.1153V}. The particles are assigned jackknife regions using a HEALPix \citep{2005ApJ...622..759G} \texttt{nside=8} tiling, as in \citet{2019MNRAS.487.2701O}, giving 169 non-empty jackknife regions each of which have equal areas on the sphere. 

In order to investigate how well we can compute the covariance matrix using solely a single dataset, the input matrix 2PCF $\hat\xi_a$ is also computed from the QPM mock, unlike in previous approaches \citep{2016MNRAS.462.2681O,2019MNRAS.487.2701O}. We use 2PCF bins of $\Delta r = 2\,h^{-1}\,$Mpc, $\Delta\mu = 0.05$ for $r\in[0,180]\,h^{-1}$\,Mpc, using narrower bins than for the covariance matrix to better capture small-scale behavior and `finger-of-god' effects.\footnote{In fact, the finer binning was found to have limited impact on the covariance matrix output, with this and the canonically-binned-$\xi$ matrices having differences corresponding to $n_\mathrm{eff}\sim 10^5$, consistent with noise. The full dependence of the covariance matrix on the input 2PCF will be discussed in future work.} $\hat\xi_a$ is computed as in Sec.\,\ref{subsec: pair_counting}, using the Landy-Szalay estimator with \texttt{corrfunc} utilized for pair counting. For the $\widetilde{DR}$ counts, the full random catalog is used, but we restrict to a randomly subsampled $N_r=10N_g$ set of random particles for the $\widetilde{RR}$ counts for computational efficiency.\footnote{This is permissible since $N_r\gg N_g$, which ensures that the shot noise error is subdominant in this term, even when using $N_r=10N_g$ randoms.} In addition, we compute 2PCF estimates for each jackknife $\hat\xi_{aA}$ (using the covariance matrix binning strategy), which are combined to compute the data jackknife covariance matrix $\hat C_{D,ab}^J$ via Eq.\,\ref{eq: C_jackknife_estimator}.

\begin{figure}%
    \centering
    \includegraphics[width=\textwidth]{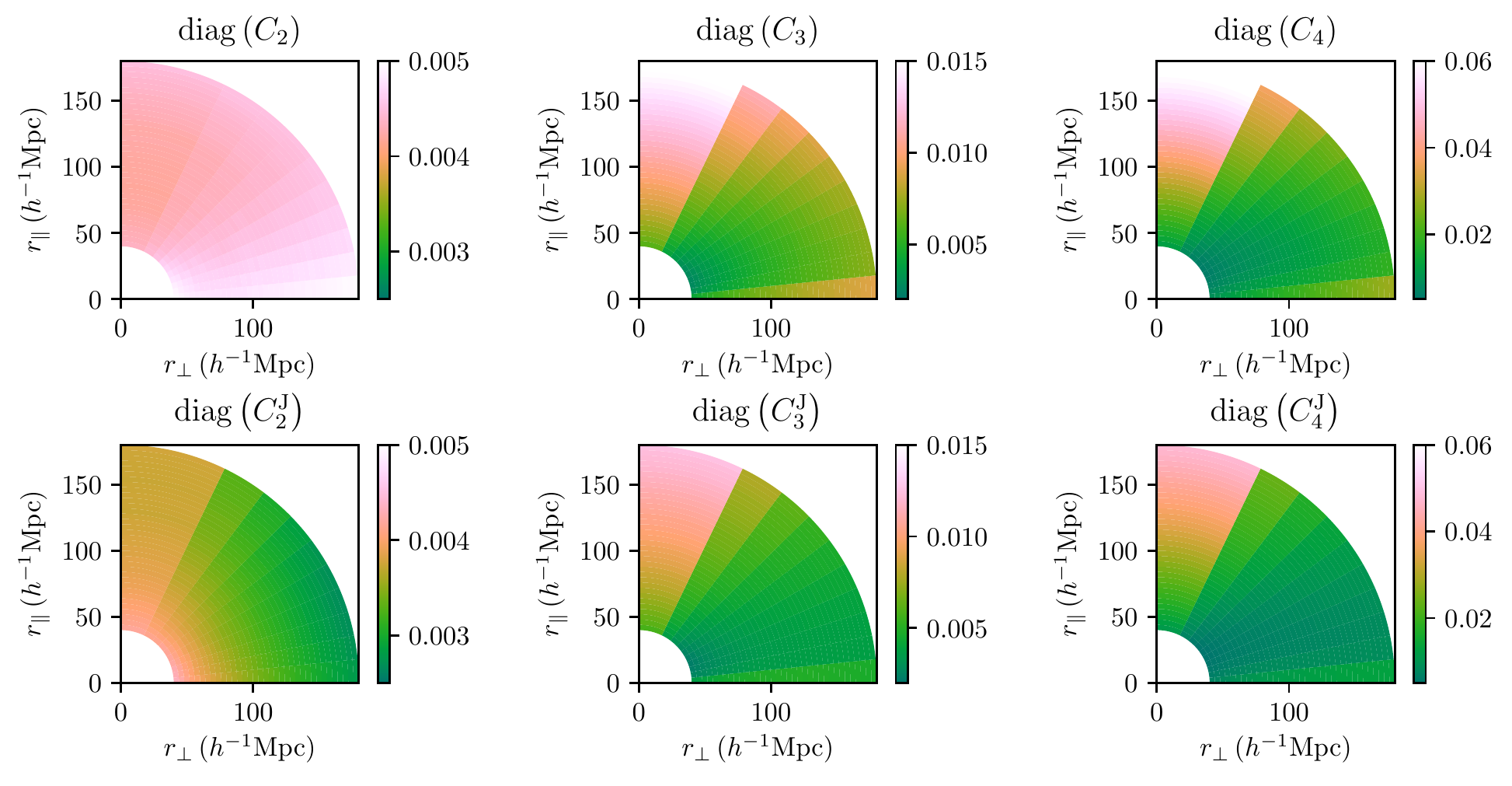}
    \caption{Diagonal elements of the 2-, 3- and 4-point full and jackknife covariance matrices, computed using 2PCF data from a single QPM mock with \texttt{RascalC}. The disconnected term, $\hat C_{x,ab}^J$ is insignificant and not shown. All covariance matrices $C_{ab}$ are multiplied by $r_ar_b$ to remove the leading scaling and are here converted into components parallel and perpendicular to the line of sight for visualization. The color scale is the same for full and jackknife terms but differs for each component, with greatest amplitudes found in the 4-point terms. The two covariance matrices are clearly similar, although we note a stronger $\mu$-dependence for the jackknife integrals. In this paper, the jackknife covariance matrices are used only for fitting a shot-noise rescaling parameter and are not good estimators of the full covariance matrices, due to violation of jackknife independence assumptions.}%
    \label{fig: diag_plot}%
\end{figure}

The covariance matrices are then estimated by running the \texttt{RascalC} code, using the $N_r=10N_g$ random particle file to define the sampling positions. Computation is performed over 20 epochs (each giving a separate estimate of $\hat C_{ab}$ and $\hat C_{ab}^J$), sampling a total of $10^{12}$ quads of particles over $\sim30$ core-hours. This gives accurate computation of the 2-, 3- and 4-point terms as well as individual matrix estimates used to compute bias-reduced estimates of the precision matrices, $\hat\Psi_{ab}$ and $\hat\Psi_{ab}^J$, via Eq.\,\ref{eq: bias_corrected_precision}. In Fig.\,\ref{fig: diag_plot}, we show the diagonal elements of the various jackknife and full covariance matrix terms computed. Notably, the combined integrals are dominated by the 4-point terms, with small contributions from the 2-point terms except on small scales. In addition, the full and jackknife matrix diagonal terms are seen to be similar, except for a stronger dependence on $\mu$ in the latter case. The similarity is as expected, since the integrals differ only by a renormalization and a $\omega_{ijkl}^{ab}$ term, which is usually close to unity.\footnote{We additionally note that the jackknife integrals appear to have smaller amplitudes than those for the full-survey; this is expected to arise from the lack of independence between jackknife regions.}

By comparing $\hat \Psi_{ab}^J(\alpha)$ and $\hat C_{D,ab}^J$ via the $\mathcal{L}_1$ likelihood (Eq.\,\ref{eq: KL_likelihood}), we may compute the optimal shot-noise rescaling parameter $\alpha$, which is found to be $\alpha^*=1.032$ here. Unlike in \citet{2019MNRAS.487.2701O}, we do not compute the error on this estimate by further jackknife computations, since individual jackknife estimates are far from independent, due to large correlations inherent in the unrestricted jackknife formalism. With this choice of $\alpha$, we obtain a noise-level corresponding to $n_\mathrm{eff}= 1.1\times 10^6$ mocks (computed via Eq.\,\ref{eq: n_eff}).

\begin{figure}%
    \centering
    \subfloat[Mock Precision]{{\includegraphics[width=0.31\textwidth]{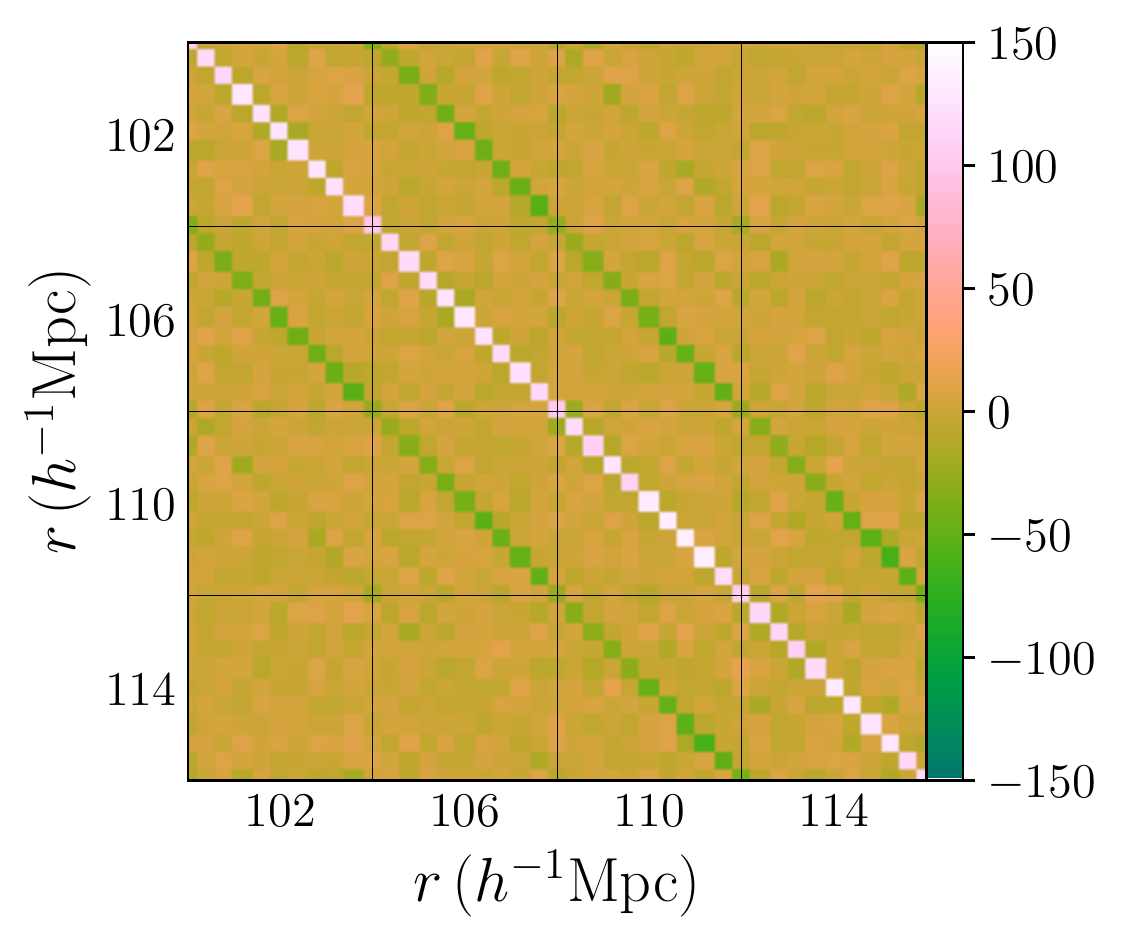} }}%
    \subfloat[Jackknife-\resub{calibrated} Model Precision]{{\includegraphics[width=0.31\textwidth]{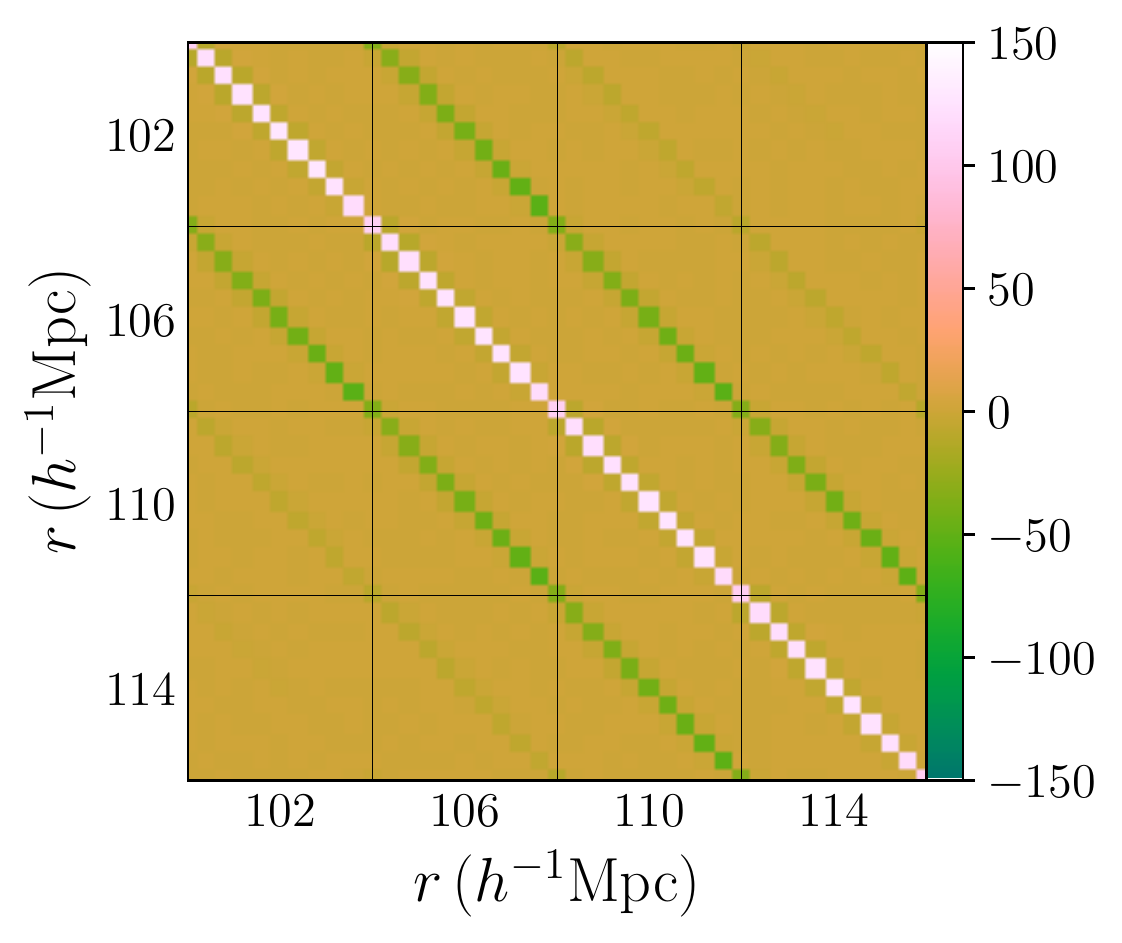} }}%
    \subfloat[(Mock - Model) Precision]{{\includegraphics[width=0.31\textwidth]{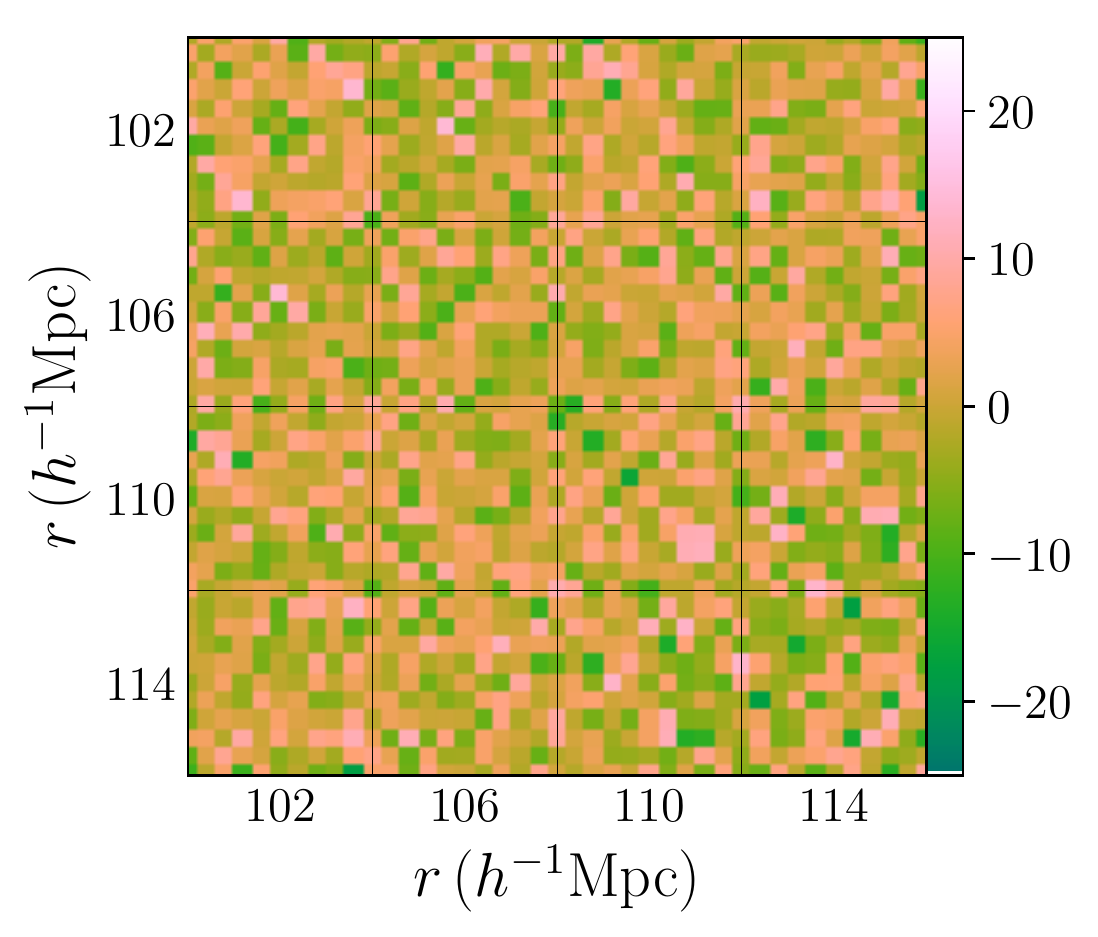} }}
    \caption{Comparison of the full-survey precision matrices, $\hat\Psi_{ab}$, (a) from 900 QPM mock catalogs and (b) using the \resub{theoretical model of Sec.\,\ref{sec: theory} which adds non-Gaussianity via a jackknife-calibrated shot-noise rescaling.} For clarity, we show only a small section of the precision matrix \resub{(with the full matrix having $r\in[40,180]\,h^{-1}\mathrm{Mpc}$)} and remove the leading scaling by dividing by $r_ar_b$. This uses \resub{narrow bins of width $\Delta\mu=0.1$, $\Delta r=4\,h^{-1}\,$Mpc (both for comparison with previous works and to stringently test our models),} with the axis labels giving the \resub{central radii of the radial bins, each of which contains 10 $\mu$ sub-bins. The model precision matrix (given by the inverse of the shot-noise-weighted sum of the full covariance matrix terms shown in Fig.\,\ref{fig: diag_plot})} uses a 2PCF derived from a single mock and a shot-noise rescaling parameter, $\alpha^{*}=1.032$, calibrated from fitting the model jackknife covariance matrix to the single-mock data. \resub{For this survey depth and binning width, the covariance is dominated by the 4-point term, rather than shot-noise, except on the smallest scales.} Plot (c) (with a zoomed-in scale) clearly shows that the mock precision matrix well represents the data.}%
    \label{fig: SingleMockQPMPlots}%
\end{figure}

\begin{figure}
\centering
\begin{minipage}[t]{.48\textwidth}
\vspace{0pt}
  \centering
  \includegraphics[width=0.95\textwidth]{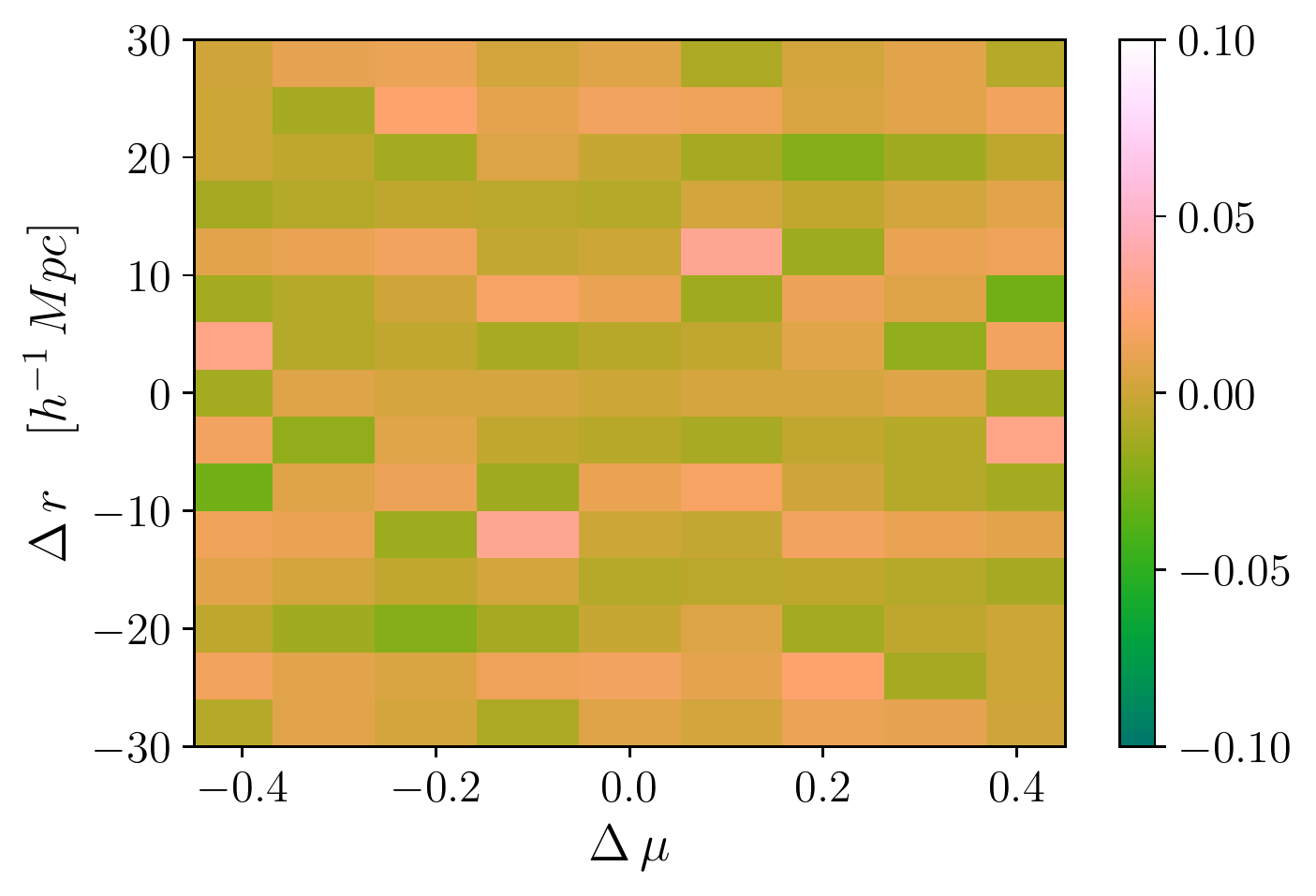}
  \caption{Stacked residuals between a 900-mock QPM precision matrix and a model estimate of $\Psi_{ab}$, with shot-noise parameter computed from calibration of model and data unrestricted jackknife covariance matrices. \resub{A section of this matrix was shown in Fig.\,\ref{fig: SingleMockQPMPlots}(c), but we now combine all elements with the same $\Delta r$ and $\Delta\mu$ values}, as in Fig.\,\ref{fig: RascalComparisonResidual}. There are no notable systematic trends in this residual matrix and it appears to be consistent with noise alone, implying that the jackknife fitting has accurately reproduced the precision matrix.}
    \label{fig: SingleMockQPMResidualStacked}
\end{minipage}%
\hfill
\begin{minipage}[t]{.48\textwidth}
\vspace{0pt}
  \centering
  \includegraphics[width=0.8\textwidth]{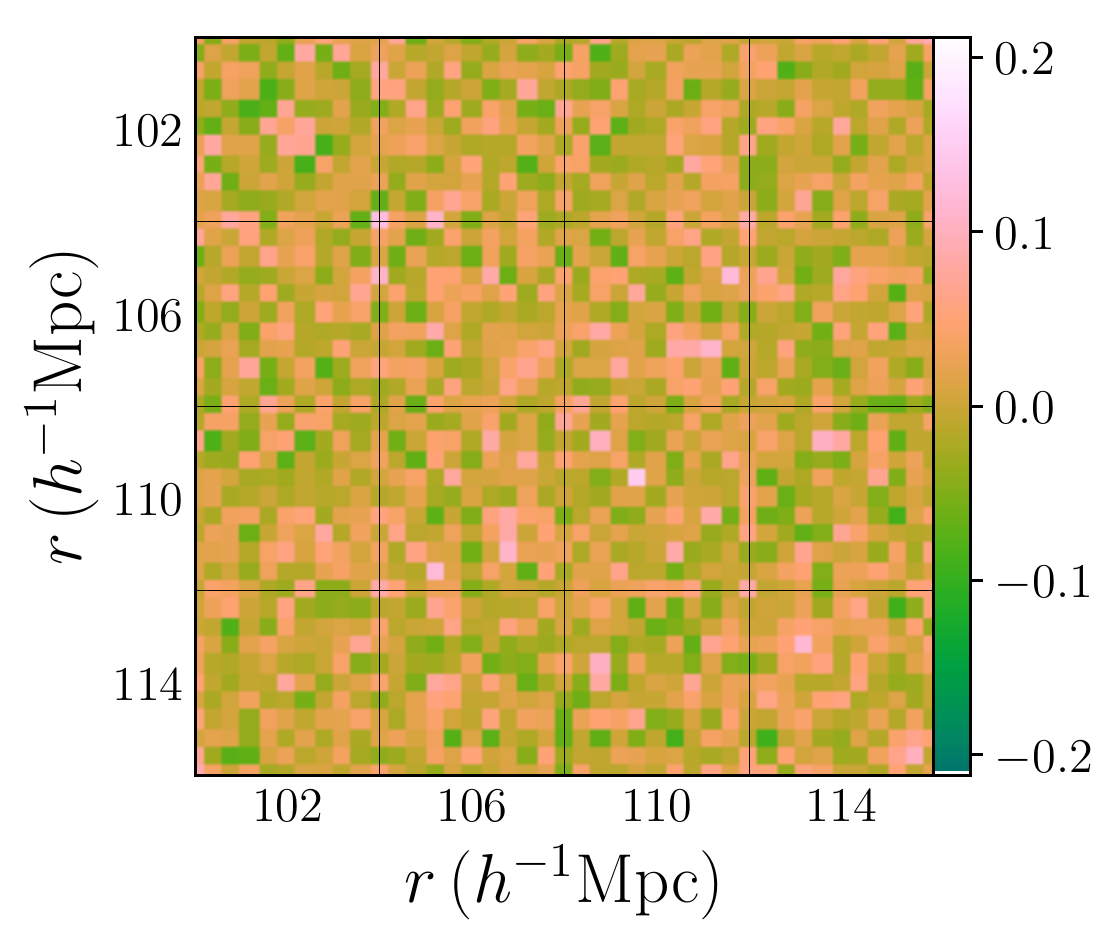}
  \caption{Section of the discriminant matrix ($\sqrt{\hat\Psi}^T\hat C_D\sqrt{\hat\Psi}-\mathbb{I}$) between a covariance matrix estimated from 900 mocks (shown inverted in Fig.\,\ref{fig: SingleMockQPMPlots}(a)) and the \texttt{RascalC} precision matrix $\hat\Psi$ (Fig.\,\ref{fig: SingleMockQPMPlots}(b)). In the limit of identical noiseless covariance matrices, this should be equal to the zero matrix. We do not observe any systematic differences from zero here, indicating that the matrices are consistent.}
  \label{fig: disc_qpm}
\end{minipage}
\end{figure}

To assess whether the derived full covariance matrices are realistic, we compare them to sample matrices from the QPM mock catalog, via the standard covariance matrix formula,
\beq\label{eq: QPM_covariance}
 \hat C_{D,ab} = \frac{1}{N_\mathrm{mocks}-1}\sum_{n=1}^{N_\mathrm{mocks}} \left[\xi_a^{(n)}-\bar\xi_a\right]\left[\xi_b^{(n)}-\bar\xi_b\right]
\eeq
where we use $N_\mathrm{mocks}=900$ QPM matrices here and exclude the mock used to compute the theoretical covariance matrix to avoid bias. A simple inversion of the $\hat C_{D,ab}$ matrix will yield a biased estimate of the QPM precision matrix; instead we use the standard form \citep{wishart28}
\beq
    \hat \Psi_{D,ab} = (1 - D)\hat C_{D,ab}^{-1}, \qquad 
    D = \frac{n_\mathrm{bins}+1}{N_\mathrm{mocks}-1}
\eeq
which we can then compare to the bias-corrected model precision matrix $\hat\Psi_{ab}(\alpha^*)$. We base our comparison on the precision matrices rather than the covariance matrices since these are more useful in later analysis (e.g. for Fisher matrix computation) and the effects of changing $\alpha$ are more clearly seen. In Fig.\,\ref{fig: SingleMockQPMPlots} we show a section of the precision matrices for the QPM mocks and the fit model as well as the residual matrix $\hat\Psi_{D,ab}-\hat\Psi_{ab}(\alpha^*)$. We note clear noise in the QPM precision matrix off-diagonal elements that is not present in the well-converged model, due to its large $n_\mathrm{eff}$. The residual matrix shows no clear trends and appears to be consistent with noise, indicating that the fitting has been done successfully. To see this more clearly, we look at the stacked residual matrix of Fig.\,\ref{fig: SingleMockQPMResidualStacked} (cf.\,Sec.\,\ref{subsec: rascal_comparison}), and note that the matrix still appears to be consistent with noise (any large deficiency along the dominant $\Psi$ elements with show up as a residual at small $\Delta r$ or $\Delta\mu$). 

A further comparison between covariance matrices is given by the discriminant matrix
\beq\label{eq: QPM_discriminant}
    \hat Q = \sqrt{\hat\Psi}^T\hat C_D\sqrt{\hat\Psi}-\mathbb{I},
\eeq
where $\sqrt{\hat\Psi}$ is the Cholesky factorization of the (jackknife-fitted) \texttt{RascalC} precision matrix, $C_D$ is the QPM covariance matrix and $\mathbb{I}$ is the identity matrix. In the absence of noise, we expect $\hat Q_{ab} = 0\,\forall(a,b)$ if the \texttt{RascalC} matrix matches that of the mocks; any systematic deviations from zero indicate differences between the two matrices. We here choose to invert the \texttt{RascalC} matrix here since it is better converged. A section of this matrix is shown in Fig.\,\ref{fig: disc_qpm}, and we observe no significant deviations from zero, with a mean value of $3\times10^{-4}$ and a standard devation of $0.02$ across all independent matrix elements. This again indicates that our model is consistent with the QPM mock covariance on large scales. 

Furthermore, we may use the $\mathcal{L}_1$ likelihood instead to compare instead the full theoretical precision matrix, $\hat\Psi(\alpha)$, and QPM covariance matrix, $\hat C_{D,ab}$, to find an optimal value for $\alpha$ without using the jackknife information, (as in \citealt{2016MNRAS.462.2681O}). This yields the estimate $\alpha^* = 1.043\pm 0.002$, where the error is given by the standard jackknife error from fitting $\hat\Psi(\alpha)$ to sets of QPM covariance mocks using only $N_\mathrm{mocks}-1$ of the $N_\mathrm{mocks}$ mocks. Notably, there is tension between this and the jackknife-derived estimate of $a$; this is expected to result from different galaxy numbers in each QPM mock since the $d$-point matrix $C_{d,ab}\propto \left(\alpha/N_g\right)^{(4-d)}$, and we observe variation of up to $3\%$ in $N_g$ across all mocks. 

Quantitatively, the matrix similarity is again assessed via the KL divergence between the two estimates, where we invert the well-converged model. This gives $D_{KL}=40.1$, which may be compared to the expected KL divergence from a noise estimate of given $N_\mathrm{mocks}$ (Eq.\,\ref{eq: relating_KL_div_to_n_samples_simple}) of $D_{KL,\mathrm{expected}}=30.7$. Note that Eq.\,\ref{eq: KL_div_from_n_eff} is an expectation value only, and strictly only true for $N_\mathrm{mocks}\gg n_\mathrm{bins}$, thus we do not expect perfect agreement here. We hence conclude that the difference between the matrices appears to be consistent with noise here.

\resub{A further test of the method would be to use the model and sample covariance matrices to compute parameter constraints (e.g. for the BAO rescaling parameters $\alpha$ and $\epsilon$) in a Fisher forecast. In \citet[Figs.\,5\,\&\,9]{2016MNRAS.462.2681O}, good agreement was found between the parameter constraints using mock data and \texttt{Rascal} model covariances (shown above to be highly consistent with those of \texttt{RascalC}), though the inclusion of non-Gaussianity in the covariances had only a minor impact on contours. For this reason, the test is not expected to be a powerful matrix discriminator and hence we do not repeat it in this analysis.}

\subsubsection{Inter-Mock Variation}
Given that we can obtain good fitting results from only a single dataset, it is worth considering the differences in the output matrices that arise from using input $\hat\xi_a$ and $\hat\xi_{aA}$ functions computed from different input mocks. To do this, we analyze a set of 20 QPM mocks in the same manner as above, computing model $\hat C_{ab}(\alpha)$ and $\hat C^J_{ab}(\alpha)$ matrices for each mock, using $N_\mathrm{quads}=10^{12}$.

As before, the theoretical jackknife covariance matrices for each mock are fit using the $\mathcal{L}_1$ likelihood to their respective $\hat C_{D,ab}^J$ matrices, giving rescaling parameters $\alpha=1.027\pm0.005$. We note that this is not directly comparable to former analyses \citep{2019MNRAS.487.2701O} using QPM mocks since each matrix is now computed with a different 2PCF input and $N_g$. The output matrices have effective mock numbers $n_\mathrm{eff}=(3.2\pm1.5)\times 10^5$, with variations arising from the different input $\hat\xi$ realizations and $N_g$.

A simple comparison between the first and $i$-th mock (for $i\in[2,20]$) is achieved via inspection of the `comparison' matrix (cf.\,Eq.\,\ref{eq: QPM_discriminant})
\beq\label{eq: discriminant}
    \hat P_{(i)} = \sqrt{\hat\Psi}_{(1)}^T\hat{C}_{(i)}\sqrt{\hat\Psi}_{(1)},
\eeq
where $\sqrt{\hat\Psi}_{(1)}$ is the Cholesky factorization of the first mock model precision matrix. In the limit $\hat C_{(i)}=\hat C_{(1)}$, this should be the identity matrix. Here, we compare full covariance matrices $\hat C_{(i)}(\alpha_i^*)$, each of which has a different $N_g$ and optimal shot-noise rescaling $\alpha^*$ so it is not \textit{a priori} known that the matrices will be similar. Fig.\,\ref{fig: disc_eigenvals} plots the distribution of eigenvalues of $\hat P_{(i)}$ for each mocks, with $\operatorname{eig}\left(\hat P_{(i)}\right)=1$ expected for identical matrices. \resub{To aid interpretation, we additionally plot the distribution of eigenvalues expected if the QPM covariance matrices are simply noisy draws from a Wishart distribution of fixed mean, i.e. the expected results for a scenario with no differences between the mocks other than those arising from noise. This computed from the average eigenspectra of 999 comparison matrices, created using 1000 draws from the Wishart distribution with mean given by that of the QPM mock covariances, setting the number of degrees of freedom to the average value of $n_\mathrm{eff}$.} Heuristically, the eigenvalues are close to unity for all mocks, \resub{with good agreement both between individual mocks and with the Wishart prediction.} We report a mean eigenvalue of $0.983 \pm 0.012$ \resub{($1.001$)} and a standard deviation of $0.056 \pm 0.005$ \resub{($0.047$)} for the mocks (Wishart prediction). \resub{These values, in tandem with Fig.\,\ref{fig: disc_eigenvals}, indicate that the output full covariance matrices do not display large variations between mocks, and are broadly consistent with the noise-only Wishart prediction, despite fair variation in the 2PCF, the number of galaxies and the noise-level (parametrized by $n_\mathrm{eff}$) between mock covariance matrices. Although this is an interesting comparison, it is important to note that non-Gaussianity is not strongly expressed in the covariance matrix eigenvalues \citep[e.g.][]{2016MNRAS.456.2662F}, with only few eigenvalues changing with the addition of non-Gaussian terms. It is thus desirable to augment our analysis with other methods.}

To systematically investigate the cause of differences between matrices, we may use the KL divergence to see if the difference is consistent with noise alone. Computing the KL divergence between all possible QPM covariance matrix pairs gives $D_{KL} = 0.35\pm0.08$, which may be compared to the $n_\mathrm{eff}$-derived expectation of $D_{KL,\mathrm{expected}} = 0.12\pm0.04$. The latter values are computed from the general KL divergence form for two matrices (Eq.\,\ref{eq: relating_KL_div_to_n_samples_general}) in appendix \ref{appen: KL-div}. Since the true KL divergence is significantly below the expected value, we conclude that there are differences between the single-mock covariance matrices that cannot be described by noise 
alone, arising from the different $N_g$ and input $\hat\xi_a$ used. Converting the measured KL-divergence to an effective number of samples via Eq.\,\ref{eq: relating_KL_div_to_n_samples_general} gives $n_{\mathrm{eff},KL} \sim 2\times 10^5$; thus this difference is not important if we require effective mock numbers $\lesssim 10^5$.

\subsection{Convergence Timescales}
Since each \texttt{RascalC} run contains a number of individual estimates of the covariance matrix estimates, we can easily assess the dependence of the effective number of mocks, $n_\mathrm{eff}$ on the number of quads sampled (and the associated run-time). Here, we consider two runs of \texttt{RascalC} over $N_\mathrm{epochs}=100$, using (a) the smoothed mean $\hat\xi_\mathrm{mean}$ from the 1000 QPM mocks (as in Sec.\,\ref{subsec: rascal_comparison}) and (b) the single mock $\hat\xi_\mathrm{single}$ estimate (as in Sec.\,\ref{subsec: qpm_cov_single}). For each run, we consider a set of subsampled covariance matrix estimates, using between 5 and 100 of the available epochs, and compute the $n_\mathrm{eff}$ for each (via the bias-correction $\tilde D$ matrix as in Sec.\,\,\ref{subsec: shot_noise_rescaling}), giving estimates of $n_\mathrm{eff}$  as a function of $N_\mathrm{epoch}$ that are converted to $N_\mathrm{quads}$ counts via the total number of quads sampled. For consistency, we use $\alpha=1$ for both matrices, noting that larger $\alpha$ gives more weight to highly converged matrices and hence larger $n_\mathrm{eff}$. 

\begin{figure}
\centering
\begin{minipage}[t]{.48\textwidth}
\vspace{0pt}
  \centering
  \includegraphics[width=0.9\textwidth]{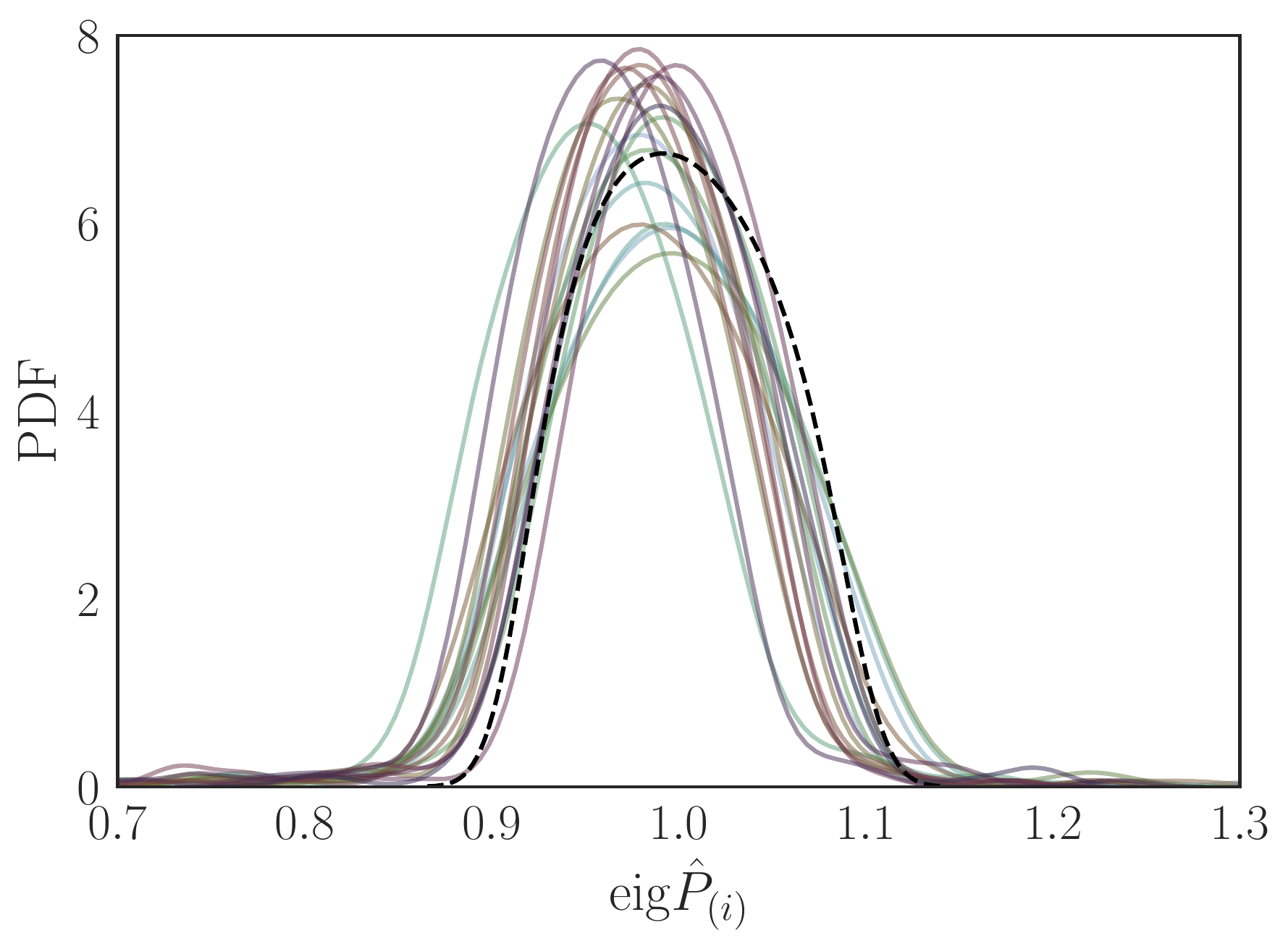}
  \caption{Distributions of the eigenvalues of the comparison matrix $\hat P_{(i)}$ between QPM mocks, as defined in Eq.\,\ref{eq: discriminant}. \resub{The dashed line shows the expected results for covariance matrices of equivalent noise drawn from a single Wishart distribution, taking the average over many such realizations}. This shows the difference between covariance matrices computed from different mocks, changing both the 2PCF and the total number of galaxies. We here display a kernel density estimate of the eigenvalue histogram, with distributions from 19 mocks overplotted (using mock 1 as a reference). For identical matrices $\hat P_{(i)}=\mathbb{I}$, thus eigenvalues should all be unity. Since the eigenvalues are here close to unity, the matrices appear to be heuristically similar in this analysis, and all mocks follow a similar trend. \resub{In addition, there is good agreement between the individual eigenspectra and the (dashed) Wishart prediction (which has non-unity eigenvalues only arising from noise), especially given that there is significant variation in the measured 2PCF and galaxy number between QPM mocks as well as in the noise levels of each mock covariance.}}
  \label{fig: disc_eigenvals}
\end{minipage}%
\hfill
\begin{minipage}[t]{.48\textwidth}
\vspace{0pt}
  \centering
  \includegraphics[width=0.9\textwidth]{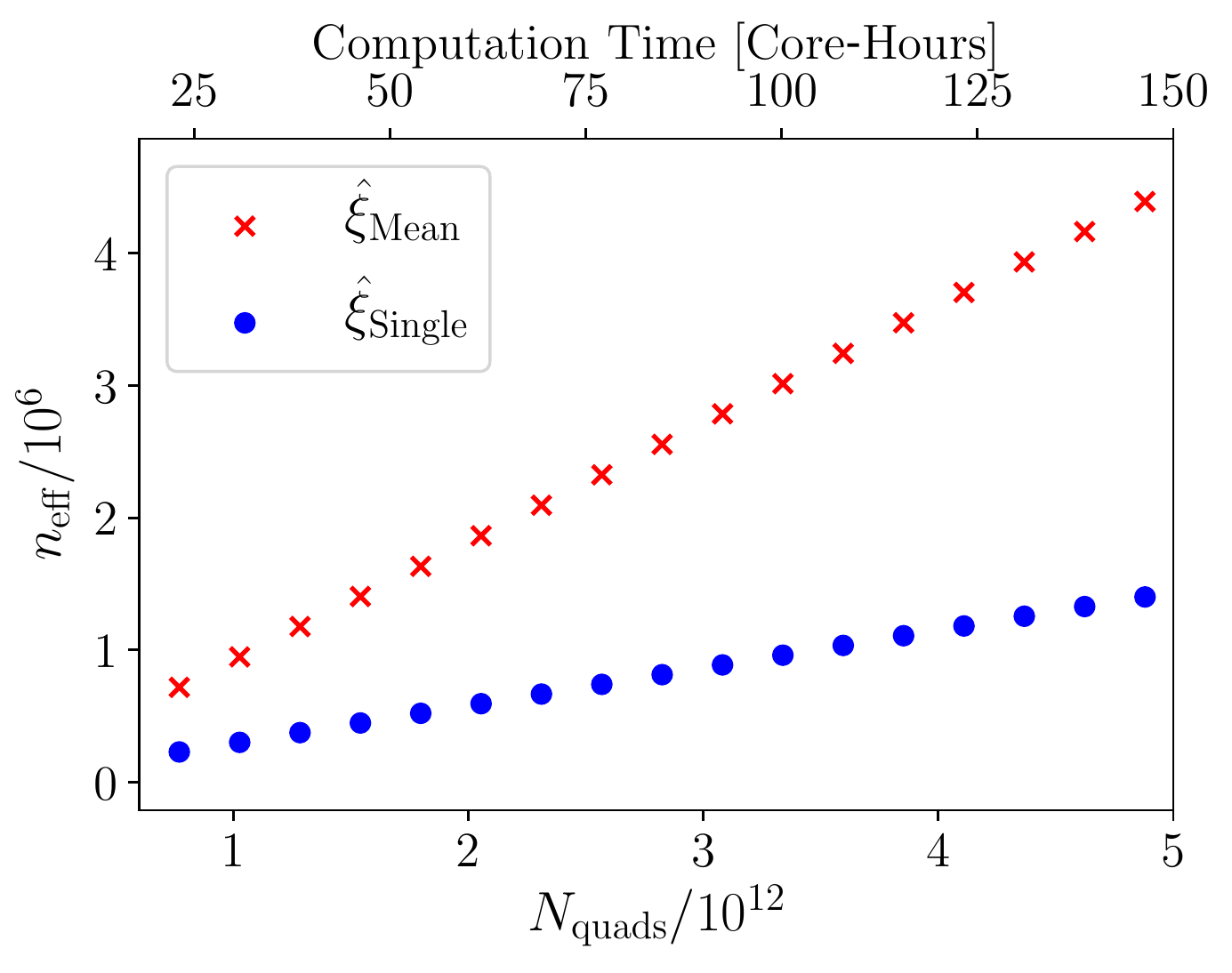}
  \caption{Dependence of the effective number of mocks, $n_\mathrm{eff}$ of the output covariance matrices on the number of quads of particles sampled, $N_\mathrm{quads}$. This is shown for two input 2PCFs: (a) $\hat\xi_\mathrm{Mean}$, a smooth 2PCF computed from the mean of 1000 QPM mocks; (b) $\hat\xi_\mathrm{Single}$, a noisy 2PCF computed from a single QPM mock, and we use shot-noise rescaling $\alpha=1$ for both. We additionally show the \texttt{RascalC} runtime on a modern CPU. $n_\mathrm{eff}$ exhibits a roughly linear relationship with $N_\mathrm{quads}$ and we note that very low noise matrices can be obtained at low computational cost. For comparison, the original \texttt{Rascal} code would require $\sim 800$ times more CPU-hours to obtain the same results.}
    \label{fig: n_eff_plot}
\end{minipage}
\end{figure}

This is shown in Fig.\,\ref{fig: n_eff_plot} for the two matrices, and we note a clear linear relationship between $N_\mathrm{quads}$ and $n_\mathrm{eff}$, implying that the noise continues to reduce as we sample the integrals more finely.  Notably, $n_\mathrm{eff}$ is larger for the smooth input 2PCF by a factor $\sim3$; we attribute this to the additional noise on the correlation function leading to a covariance matrix which takes longer to converge. Looking at the computation time for these runs, it is clear that \texttt{RascalC} is able to estimate covariance matrices at very low noise levels ($n_\mathrm{eff}\sim10^6$) in a few tens of CPU-hours.\footnote{Note that $n_\mathrm{eff}\rightarrow\infty$ does not imply that our model is correct, rather that the matrices are fully converged. Tests like those in Sec.\,\ref{subsec: qpm_cov_single} are required to assess the validity of the approach.} There is a small extra computational overhead due to the pre-computation of the $\hat\xi_{aA}$ and $w_{aA}$ functions; this takes a few tens of CPU hours, but only needs be done once for each survey geometry (for $w_{aA}$) or mock (for $\hat\xi_{aA}$). In addition, the number of quads required for convergence to a given $n_\mathrm{eff}$ level depends strongly on the covariance matrix binning; reducing the number of bins (here 350) by a factor $\beta$ gives $\beta^2$ less covariance matrix elements thus $\beta^2$ more counts per bin, leading to convergence accelerated by a factor $\sim\beta^2$. 

\section{Multiple Field Generalization}\label{sec: general_cov_matrices}
In the above sections, we have considered only the \textit{auto-covariance} for a single set of tracer galaxies; that is, the covariance of the 2PCF $\xi$ with itself. Here, we generalize the above formalism to compute \textit{cross-covariances} between two correlation functions $\xi^{XY}$ and $\xi^{ZW}$, where the labels $X,Y,Z,W$ refer to (possibly distinct) sets of tracer particles. This is most important for the two field case (with fields labelled $S$ and $T$), where we can compute the covariances between any combinations of the 2PCFs $\{\xi^{SS},\xi^{ST},\xi^{TT}\}$. This has applications for upcoming surveys, for example cross-correlating the ELG and LRG populations in eBOSS \citep{2016AJ....151...44D} or DESI \citep{2016arXiv161100036D}. An application to data will be given in future work.

\subsection{Generalized Cross-Covariance Matrices and Non-Gaussianity}
As before, we start with the 2PCF definitions for single jackknifes;
\beq\label{eq: RR_aAXY}
    \hat{\xi}_{aA}^{XY} &=& \frac{1}{RR_{aA}^{XY}}\sum_{i\neq{}j}n_i^Xn_j^Yw_i^Xw_j^Y\Theta_a^{ij}q_{ij}^A\delta_i^X\delta_j^Y\\\nonumber
    RR_{aA}^{XY} &=& \sum_{i\neq{}j}n_i^Xn_j^Yw_i^Xw_j^Y\Theta_a^{ij}q_{ij}^A,
\eeq
and the summed estimates
\beq\label{eq: RR_aXY}
    \hat{\xi}_{a,J}^{XY} &=& \sum_Aw_{aA}^{XY}\xi_{aA}^{XY} = \frac{1}{RR_{a,J}^{XY}}\sum_{i\neq{}j}n_i^Xn_j^Yw_i^Xw_j^Y\Theta_a^{ij}Q_{ij}\delta_i^X\delta_j^Y\\\nonumber
    w_{aA}^{XY} &=& \frac{RR_{aA}^{XY}}{RR_{a,J}^{XY}}\\\nonumber
    RR_{a,J}^{XY} &=& \sum_ARR_{aA}^{XY} = \sum_{i\neq{}j}n_i^Xn_j^Yw_i^Xw_j^Y\Theta_a^{ij}Q_{ij}
\eeq
using superscripts to label the relevant fields. $q_{ij}^A$ and $Q_{ij}$ have the same definitions as before. In the case of the unrestricted jackknife, $\hat \xi_{a,J}^{XY}$ and $RR_{a,J}^{XY}$ are equal to the respective full-survey forms ($\hat\xi_a^{XY}$ and $RR_a^{XY}$) that would be obtained in the absence of jackknifes, as before. The standard full and jackknife covariance matrices are transformed to the 4-field form
\beq\label{eq: generalized_covariance_form}
    {C}_{ab}^{XY,ZW} &=& \av{\hat\xi^{XY}_a\hat\xi^{ZW}_b}-\av{\hat\xi^{XY}_a}\av{\hat\xi^{ZW}_b}\\\nonumber
    {C}_{ab,J}^{XY,ZW} &=& \frac{1}{1-\sum_Bw_{aB}^{XY}w_{bB}^{ZW}}\sum_Aw_{aA}^{XY}w_{bA}^{ZW}\left(\hat{\xi}_{aA}^{XY}-\hat{\xi}_{a,J}^{XY}\right)\left(\hat{\xi}_{bA}^{ZW}-\hat{\xi}_{b,J}^{ZW}\right),
\eeq
which can be expanded into 2-, 3- and 4-point terms as before. A key difference between this and the auto-covariance case is that we can only contract \textit{identical} fields using the shot-noise identity, i.e. the expression
\beq
    \delta_i^X\delta_i^Y \approx \delta^{XY}\frac{\alpha^X}{n_i^X}\left(1+\delta^X_i\right)
\eeq
now includes a Kronecker delta $\delta^{XY}$, with shot-noise rescaling parameter for the $X$ field $\alpha^X$. This implies that we only have 2- and 3-point contributions for certain field combinations; a 3-point term requires at least an identical pair of fields between the first and second correlation input function and the 2-point term needs both 2PCFs to be identical. This therefore limits the amount of non-Gaussianity that can be encapsulated by our simple shot-noise rescaling parameter. Our methodology, however, is flexible, and could be extended to incorporate some method of coupling non-identical fields.

Inserting the cross-correlation function definitions (Eqs.\,\ref{eq: RR_aAXY}\,\&\,\ref{eq: RR_aXY}) into these covariances, we can derive the following expansion for the covariance matrices, which applies to both full and jackknife matrices;
\beq\label{eq: theoretical_generalized_jackknife_matrix}
    \av{C_{ab}^{XY,ZW}} &=& {C}_{4,ab}^{XY,ZW} + \frac{\alpha^X}{4} \left[\delta^{XW}{C}_{3,ab}^{X,YZ}+\delta^{XZ}{C}_{3,ab}^{X,YW}\right]+\frac{\alpha^Y}{4}\left[\delta^{YW}{C}_{3,ab}^{Y,XZ}+\delta^{YZ}{C}_{3,ab}^{Y,XW}\right]\\\nonumber
    &+&\frac{\alpha^X\alpha^Y}{2}\left(\delta^{XW}\delta^{YZ}+\delta^{XZ}\delta^{YW}\right){C}_{2,ab}^{XY}.
\eeq
This uses the following definitions of the jackknife covariance matrices (contracting over $j$ in the 3-point term as before):
\beq\label{eq: generalized_covariance_definitions}
    {C}_{4,ab,J}^{XY,ZW} &=& \frac{1}{1-\sum_Bw_{aB}^{XY}w_{bB}^{ZW}}\frac{1}{RR_a^{XY}RR_b^{ZW}}\sum_{i\neq{}j\neq{}k\neq{}l}n_i^Xn_j^Yn_k^Zn_l^Ww_i^Xw_j^Yw_k^Zw_l^W\Theta_a^{ij}\Theta_b^{kl}\omega_{ijkl,ab}^{XY,ZW}\\\nonumber
    &\times& \left(\xi^{XYZW}_{ijkl}+\xi_{ik}^{XZ}\xi_{jl}^{YW}+\xi_{il}^{XW}\xi_{jk}^{YZ}+\xi_{ij}^{XY}\xi_{kl}^{ZW}\right)\\\nonumber
    {C}_{3,ab,J}^{Y,XZ} &=& \frac{4}{1-\sum_Bw_{aB}^{XY}w_{bB}^{ZY}}\frac{1}{RR_a^{XY}RR_b^{ZY}} \sum_{i\neq{}j\neq{}k}n_i^Xn_j^Yn_k^Zw_i^X\left(w_j^Y\right)^2w_k^Z\Theta^{ij}_a\Theta^{jk}_b\omega_{ijkj,ab}^{XY,ZY}\left(\xi_{ijk}^{XYZ}+\xi_{ik}^{XZ}\right)\\\nonumber
    {C}_{2,ab,J}^{XY} &=& \frac{2\delta^{ab}}{1-\sum_Bw_{aB}^{XY}w_{bB}^{XY}}\frac{1}{RR_a^{XY}RR_b^{XY}}\sum_{i\neq{}j}n_i^Xn_j^Y\left(w_i^X\right)^2\left(w_j^Y\right)^2\Theta_a^{ij}\omega_{ijij,ab}^{XY,XY}\left(1+\xi_{ij}^{XY}\right)
\eeq
using the generalized weighting tensor
\beq\label{eq: jackknife_weight_tensor_generalized}
    \omega_{ijkl,ab}^{XY,ZW} = \sum_A\left(q_{ij}^A-w_{aA}^{XY}Q_{ij}\right)\left(q_{kl}^A-w_{bA}^{ZW}Q_{kl}\right).
\eeq
In Eqs.\,\ref{eq: generalized_covariance_definitions}, we include the non-Gaussian terms for completeness (with $\xi$ terms involving 3- and 4- indices understood to be 3- and 4-point correlation functions) and insert symmetry factors for the 3- and 4-point integrals for compatibility with the single-field forms. These reduce to the forms listed above in the case $X=Y=Z=W$, and we note that there is still a disconnect term in the 4-point function, which should be separately computed, as in Sec.\,\ref{subsec: disconnected_evaluation}. The non-jackknife matrices have a similar form, excluding the normalizing prefactor, the $\omega_{ijkl}$ tensor and the disconnected term.

These expression have a number of symmetries with respect to field interchanges. First, we note that $\omega_{ijkl,ab}^{XY,ZW}$ is invariant under $i\leftrightarrow j$, $k\leftrightarrow l$, $X\leftrightarrow Y$, $Z\leftrightarrow W$, and that $\xi^{XY}_{ij}$ is symmetric in the two fields $X$ and $Y$. This implies both that $C_{4,ab,J}^{XY,ZW}$ is invariant under the transformations $X\leftrightarrow{}Y$ or $Z\leftrightarrow{}W$. In addition, we have the transpose-symmetry $C_{4,ab,J}^{XY,ZW}=C_{4,ba,J}^{ZW,XY}=\left(C_{4,ab,J}^{ZW,XY}\right)^T$. For the 3-point terms, we have similar symmetry in the final two indices; $C_{3,ab,J}^{X,YZ}=C_{3,ba,J}^{X,ZY} = \left(C_{3,ab,J}^{X,ZY}\right)^T$. In the single-field case, we were able to replace the 4-point $\left(\xi_{ik}\xi_{jl}+\xi_{il}\xi_{jk}\right)$ term with $2\xi_{ik}\xi_{jl}$ for computational efficiency. Here, this is true for all two-field terms except the $C_{4,ab}^{ST,ST}$ term and its symmetries (for fields $S$ and $T$). To aid computation, we will assume this simplification, but compute both $C_{4,ab}^{ST,ST}$ and $C_{4,ab}^{ST,TS}$, which, when averaged together, give the correct 4-point term.

\subsection{Generalized Correlation Function Estimators}
The generalized 2PCFs, used to compute both cross-covariances and the shot-noise parameter, follow a similar definition to the single-field case (cf.\,Sec.\,\ref{subsec: corr_functions}). For general fields $X$ and $Y$, the full-survey 2PCF may be estimated by the symmetric \citet{1993ApJ...412...64L} generalization
\beq\label{eq: cross_correlation_LS_multi}
    \hat{\xi}^{XY}_{a} = \frac{\widetilde{DD}^{XY}_{a} - \widetilde{DR}^{XY}_{a} - \widetilde{DR}^{YX}_{a}+\widetilde{RR}^{XY}_{a}}{\widetilde{RR}^{XY}_{a}}
\eeq
where $\widetilde{DR}^{XY}$ indicates pair counts using the $X$ data and $Y$ random fields for example. These pair counts have the same form as for the single field case (Eq.\,\ref{eq: weighted_pair_count}), with fields $F,G\in\{D^X,D^Y,R^X,R^Y\}$. In practice, computing the set of $\{\xi^{XY}\}$ this simply involves computing pair counts for all combinations of $D$ and $R$ fields. For unrestricted jackknife 2PCFs, we have a similar form
\begin{eqnarray}\label{eq: jackknife_correlation_LS_multi}
    \hat{\xi}_{aA}^{J,XY} &=& \frac{\widetilde{DD}^{XY}_{aA} - \widetilde{DR}^{XY}_{aA} - \widetilde{DR}^{YX}_{aA}+\widetilde{RR}^{XY}_{aA}}{\widetilde{RR}^{XY}_{aA}}
\end{eqnarray}
with jackknife pair counts defined as in Eq.\,\ref{eq: weighted_pair_count_jk} and computed as before e.g. we compute the $\widetilde{D^XR^Y}_{aA}$ pair counts by taking the mean of \texttt{corrfunc} pair counts of (a) the entire $D^X$ field with the $A$ jackknife of $R^Y$ and (b) the entire $R^Y$ field with the $A$ jackknife of $D^X$. These are then used to compute the data jackknife cross-covariance matrix via Eq.\,\ref{eq: generalized_covariance_form}.

\subsection{Computing the Gaussian Generalized Covariances}
We now specialize to the 2-field case, with distinct fields labelled $S$ and $T$ (which could represent different galaxy populations), and neglect any non-Gaussian components, except for the shot-noise rescaling. In this case, there are six independent covariance matrices, yet each can depend on a number of 2- and 3-point terms. The above symmetry constraints limit the total number of submatrices that need be computed to three 2-point terms, six 3-point terms and six 4-point terms. 

We may use the algorithm described in Sec.\,\ref{sec: algorithm_design} to compute the relevant matrices with minimal alterations. Instead of assuming the $i$, $j$, $k$ and $l$ fields to be identical, we now draw them from $X$, $Y$, $Z$ and $W$ fields respectively (with $X,Y,Z,W\in\{S,T\}$), and the algorithm computes a single 2-, 3- and 4-point (full and jackknife) matrix. For efficient importance sampling, it is clear from the form of $C_{4,ab}^{XY,ZW}$ that we must choose cell $c_j$ from cell $c_i$ via a $1/r^2$ kernel as before, yet now choose $c_k$ from $c_i$ via a $\bar\xi^{XZ}(r)$ kernel and $c_l$ from $c_j$ via a $\bar\xi^{YW}(r)$ kernel. Due to the ordering of fields in the 3-point term $C_{3,ab}^{Y,XZ}$, this also samples the 3-point term effectively.

If these fields are ordered correctly, we may compute all required covariance matrix terms (for the six non-trivial cross-covariances) in only six iterations of the algorithm, whilst retaining the desired precision boosts from importance sampling. As mentioned above, we assume the simplification $\left(\xi_{ik}^{XZ}\xi_{jl}^{YW}+\xi_{il}^{XW}\xi_{jk}^{YZ}\right) = 2\xi_{ik}^{XZ}\xi_{jl}^{YW}$ in the 4-point integrals, which requires us to compute both $C_{4,ab}^{ST,TS}$ and $C_{4,ab}^{ST,TS}$ to correctly estimate $C_{4,ab}^{ST,ST}$, giving a total of seven runs of the algorithm. An example of optimal field ordering is shown in Tab.\,\ref{tab: generalized_matrix_ordering}, listing the desired $i$, $j$, $k$ and $l$ fields and the associated terms computed.

\begin{table}
\centering
\caption{An example of optimal field ordering to compute all cross-covariance matrix terms for two random fields $S$ and $T$. The first four columns indicate the fields from which $i$, $j$, $k$ and $l$ particles are drawn and the subsequent columns indicate the associated covariance submatrices computed. The 2- and 3-point terms in brackets indicate quantities that have already been computed. We compute both $(ST,TS)$ and $(ST,ST)$ 4-point terms to allow us to use a simplified 4-point estimator for more efficient importance sampling, as described in the text.}\label{tab: generalized_matrix_ordering}
\begin{tabular}{cccc|c|c|c|}
\hline
\multicolumn{4}{c}{Fields} & \multicolumn{3}{c}{Submatrices}\\
\cmidrule(lr){1-4}\cmidrule(lr){5-7}
$i$ & $j$ & $k$ & $l$ & 4-point & 3-point & 2-point \\ \hline
S&S&S&S & SS,SS   & S,SS    & SS      \\
S&T&S&S & ST,SS   & T,SS    & ST      \\
S&T&T&S & ST,TS   & T,ST    & (ST)    \\
S&T&S&T & ST,ST   & (T,SS) & (ST)     \\
S&S&T&T & SS,TT   & S,ST    & (SS)    \\
T&S&T&T & TS,TT   & S,TT    & (ST)    \\
T&T&T&T &TT,TT   & T,TT    & TT     
\end{tabular}
\end{table}

\subsection{Estimating Shot Noise Parameters}
Following computation of the submatrices above, the matrices can be reconstructed using their symmetry properties. To compute the shot-noise rescaling parameters, we must compare the theoretical cross-covariance matrices with those derived from estimates of the jackknife 2PCFs. For the two field case, there are two rescaling parameters; $\alpha^S$ and $\alpha^T$, which are constrained by five non-trivial data-derived covariance matrices: $\{\hat C_{D,J}^{SS,SS}, \hat C_{D,J}^{SS,ST}, \hat C_{D,J}^{ST,ST}, \hat C_{D,J}^{ST,TT}, \hat C_{D,J}^{TT,TT}\}$ (excluding $\hat C_{D,J}^{SS,TT}$ which is independent of $\alpha^S$ and $\alpha^T$). Since we do not expect full correlations between the fields (i.e. $\xi^{ST}/\sqrt{\xi^{SS}\xi^{TT}}<1$), the constraints from cross-correlation terms are expected to be subdominant, thus we simply compute $\alpha^S$ and $\alpha^T$ using the KL divergences of $\hat C_{D,J}^{SS,SS}$ and $\hat C_{D,J}^{TT,TT}$ independently. This gives the estimates
\beq\label{eq: generalized_shot_noise}
\hat\alpha^S &=& \operatorname*{arg\,min}_{\alpha^S}\left[D_{KL}\left(\hat\Psi_J^{SS,SS}(\alpha),\hat C_{D,J}^{SS,SS}\right)\right]\\\nonumber
    \hat\alpha^T &=& \operatorname*{arg\,min}_{\alpha^T}\left[D_{KL}\left(\hat \Psi_J^{TT,TT}(\alpha),\hat C_{D,J}^{TT,TT}\right)\right],
\eeq
using the single-field $D_{KL}$ definition (Eq.\,\ref{eq: KL_likelihood}). This can be used to compute the final estimates of the full covariance matrices, as for the single field case.

\section{Outlook}\label{sec: conclusions}
In this paper we have outlined a new algorithm for generating model covariance matrices for galaxy 2PCFs purely from a single dataset, without reference to mocks. \resub{By using new importance sampling techniques coupled with random particle catalogs, our algorithm is able to compute covariance matrices $\sim 10^4$ times faster than previous codes \citep{2016MNRAS.462.2681O,2019MNRAS.487.2701O} with no loss of accuracy. Finely binned matrices with negligible sampling noise can be computed in less than $100$ CPU-hours; a vast improvement over mock-based approaches. We include non-Gaussianity via a slight enhancement of small-scale shot-noise power, which is found to be a good approximation for BAO-scale analyses. The rescaling can calibrated from the same data-set, by fitting a theoretical jackknife model (which depends on the same shot-noise parameter) to a sample jackknife covariance drawn from the observed data.} In addition, we have discussed the theory for model cross-covariance matrix computation, which will be of great importance in future cosmology surveys. Our fast and flexible analysis code, \texttt{RascalC}, has been made publicly available,\footnote{\url{https://RascalC.readthedocs.io}} allowing computation of fitted covariance matrix models with only galaxy position and random particle catalogs as inputs. \resub{This additionally takes care of all necessary pre- and post-processing steps (such as 2PCF estimation and jackknife fitting) and can be simply extended to more complex scenarios, e.g. three-point correlation function covariances \citep{3pcfCov}.}

Using only data from a single mock (with no prior knowledge of the 2PCF), \texttt{RascalC} was shown to produce an output precision matrix for the large-scale 2PCF that was fully consistent to that of a suite of mocks, within sampling noise. In addition, variations in the precision matrix from using different input mocks were found to only be important at noise levels corresponding to $\sim10^5$ mocks, indicating that the matrices are insensitive to the exact correlation function estimate. Although mock galaxy catalogs remain crucial for testing features such as systematic uncertainties, we hope that procedures such as this will reduce the total number of mocks required, allowing more computational power to be invested instead in their accuracy.

Having the ability to generate covariance matrices in a matter of hours will open up a range of topics for exploration; the dependence of the covariance matrix on different aspects of the 2PCF is one such example. It now becomes simple to see how the covariance matrix changes as a result of different cosmologies, without having to compute new simulations at high computational cost. Furthermore, application of the generalized cross-covariance model to upcoming multi-tracer survey data (for example from eBOSS and DESI) will allow Fisher matrices to be computed for any combination of auto- and cross-correlation functions, increasing the utility of such data.

Throughout this paper we have considered only a simple model of the covariance matrix, with non-Gaussianities simply included via a shot-noise rescaling. Notwithstanding, the full-survey and jackknife covariance integrals presented above are fully general and can be applied to any models of the connected 3- and 4-point functions, including those with a number of free parameters. Any such model may be simply computed via minor modifications to our main algorithm, with additional parameters calibrated using jackknife data in post-processing. An extension of this model to the covariances of two- and three-point functions in Legendre multipole space is presented in our accompanying work \citep{3pcfCov}. The \texttt{RascalC} code is thus applicable to a wide range of analyses, allowing for precise and accurate covariances to be computed in a fraction of the previous computational time.

\section*{Acknowledgements}

\resub{We thank the anonymous referee for suggesting useful references and helping to improve the clarity of this paper.} OHEP acknowledges funding from the Herchel-Smith foundation. DJE is supported by U.S. Department of Energy grant DE-SC0013718 and as a Simons Foundation Investigator. AW acknowledges support from the German research organization
DFG, Grants No.~WI 4501/1--1 and No.~WI 4501/2--1.

Funding for SDSS-III has been provided by the Alfred P. Sloan Foundation, the Participating Institutions, the National Science Foundation, and the U.S. Department of Energy Office of Science. The SDSS-III web site is http://www.sdss3.org/.

SDSS-III is managed by the Astrophysical Research Consortium for the Participating Institutions of the SDSS-III Collaboration including the University of Arizona, the Brazilian Participation Group, Brookhaven National Laboratory, Carnegie Mellon University, University of Florida, the French Participation Group, the German Participation Group, Harvard University, the Instituto de Astrofisica de Canarias, the Michigan State/Notre Dame/JINA Participation Group, Johns Hopkins University, Lawrence Berkeley National Laboratory, Max Planck Institute for Astrophysics, Max Planck Institute for Extraterrestrial Physics, New Mexico State University, New York University, Ohio State University, Pennsylvania State University, University of Portsmouth, Princeton University, the Spanish Participation Group, University of Tokyo, University of Utah, Vanderbilt University, University of Virginia, University of Washington, and Yale University. 



\bibliographystyle{mnras}
\bibliography{bibliography,bibliography_alex} 



\appendix
\section{Derivation of the Jackknife Covariance Matrix Integrals}\label{appen: jackknife_expansion}
To compute the expectation of the jackknife covariance matrix $\hat C_{ab}^J$ we first insert the expansions of $\hat\xi_{aA}$ and $\hat\xi_a^J$ (Eqs.\,\ref{eq: single_jackknife_xi}\,\&\,\ref{eq: xiJ_estimator_expanded}). Noting that
\beq
\bar{w}_{aA}\left(\hat{\xi}_{aA}-\hat{\xi}_a^J\right) &=& 
    \frac{1}{RR_a^J}\sum_{i\neq{}j}\left(q_{ij}^A - w_{aA}Q_{ij}\right)D_a^{ij}
\eeq
(using the definition of $\bar{w}_{aA}$ and summing over small cells $i$ and $j$), we arrive at the simplified form
\beq\label{eq: expanded_covariance_matrix}
\hat{C}_{ab}^J &=& \frac{1}{1-\sum_B w_{bA}w_{bB}}\frac{1}{RR_a^JRR_b^J}\left[\sum_{i\neq j}\sum_{k\neq{}l}D_a^{ij}D_b^{kl}\sum_A\left(q_{ij}^A-w_{aA}Q_{ij}\right)\left(q_{kl}^A-w_{bA}Q_{kl}\right)\right]\\\nonumber
&=&\frac{1}{RR_a^JRR_b^J-\sum_BRR_{aB}RR_{bB}}\left[\sum_{i\neq{}j}\sum_{k\neq{}l}D_a^{ij}D_b^{kl}\omega_{ijkl}^{ab}\right]
\eeq
using the weighting tensor $\omega_{ijkl}^{ab}$ (Eq.\,\ref{eq: omega_tensor_definition}).

The theoretical covariance matrix is given by the expectation of $\hat C_{ab}^J$, including the term  $\langle{\left(\hat{\xi}_{aA}-\hat{\xi}_a^J\right)\left(\hat{\xi}_{bA}-\hat{\xi}_b^J\right)\rangle}$ with an expectation over both $\xi_a^J$ and $\xi_{aA}$ estimators. This could alternatively be computed as $\langle{\left(\hat{\xi}_{aA}-\langle{\hat{\xi}_a^J\rangle}\right)\left(\hat{\xi}_{bA}-\langle{\hat{\xi}_b^J\rangle}\right)\rangle}$ (i.e. taking the expectation of $\xi^J_a$ before computing the expectation of the product). The two formalisms differ in the appearance of a disconnected term (cf.\,Sec.\,\ref{subsec: disconnected_term}). In our case, since we compute $\xi_{aA}$ and $\xi^J_a$ from the same dataset, it is important to use the former approach. The latter case would be appropriate when the mean 2PCF was instead estimated from independent mocks. 

The summation of Eq.\,\ref{eq: expanded_covariance_matrix} includes a sum over $i\neq j$ and $k \neq l$; this can be recast into a series of 2-, 3- and 4-point terms via
\beq\label{eq: n_point_expansion}
    \hat C_{ab}^J = \frac{1}{RR_a^JRR_b^J-\sum_BRR_{aB}RR_{bB}}\left\{\sum_{i\neq j\neq k\neq l} D_a^{ij}D_b^{kl}\omega_{ijkl}^{ab}+4\sum_{i\neq j\neq k}D_a^{ij}D_b^{jk}\omega_{ijjk}^{ab}+2\delta_{ab}\sum_{i\neq j}D_a^{ij}D_b^{ij}\omega_{ijij}^{ab}\right\}
\eeq 
for Kronecker delta $\delta_{ab}$, noting that $\omega_{ijkl}^{ab}$ is symmetric under the interchanges $i\leftrightarrow{}j$, $k\leftrightarrow{}l$ or $(i,j,a)\leftrightarrow(k,l,b)$ and $D_a^{ij}=D_a^{ji}$. 

The expectation $\av{C_{ab}^J}$ will include terms involving the expectation of two $D_a^{ij}$ factors, themselves depending on expectations of four overdensity fields e.g. $\av{\delta_i\delta_j\delta_j\delta_k}$ for the 3-point term. Contracted terms may be simplified using the shot-noise identity
\beq
     \delta_i^2 \approx \frac{1}{n_i}(1+\delta_i),
\eeq
true at leading order for small cells $i$ containing at most one particle. Using Isserlis' (Wick's) theorem \citep{isserlis} we define
\beq\label{eq: Wick expansion}
    \av{\delta_i\delta_j} &=& \xi_{ij}\\\nonumber
    \av{\delta_i\delta_j\delta_k} &=& \xi^{(3)}_{ijk}\\\nonumber
    \av{\delta_i\delta_j\delta_k\delta_l} &=& \xi^{(4)}_{ijkl}+\xi_{ij}\xi_{kl}+\xi_{ik}\xi_{jl}+\xi_{il}\xi_{jk}
\eeq
where $\xi^{(n)}$ represents the connected $n$-point correlation function (at positions defined by subscripts e.g. $\xi^{(3)}_{ijk} = \xi^{(3)}(\vec{x}_i,\vec{x}_j\vec{x}_k)$). This gives the following simplifications;
\beq\label{eq: random_field_expansions}
    \av{D_a^{ij}D_b^{kl}} &=& R_a^{ij}R_b^{kl}\av{\delta_i\delta_j\delta_k\delta_l} = R_a^{ij}R_b^{kl}\left(\xi^{(4)}_{ijkl}+\xi_{ij}\xi_{kl}+\xi_{ik}\xi_{jl}+\xi_{il}\xi_{jk}\right)\\\nonumber
    \av{D_a^{ij}D_b^{jk}} &=& \frac{1}{n_j}R_a^{ij}R_b^{jk}\av{\delta_i\delta_k(1+\delta_j)} = \frac{1}{n_j}R_a^{ij}R_b^{jk}\left(\xi^{(3)}_{ijk}+\xi_{ik}\right)\\\nonumber
    \av{D_a^{ij}D_b^{ij}} &=& \frac{1}{n_in_j}R_a^{ij}R_b^{ij}\av{(1+\delta_i)(1+\delta_j)} = \frac{1}{n_in_j}R_a^{ij}R_b^{ij}\left(1+\xi_{ij}\right)
\eeq

In this paper, we choose to model non-Gaussianity via a \textit{shot-noise rescaling}, a technique whose effectiveness on large-scales was clearly demonstrated in \citet{2016MNRAS.462.2681O} and \citet{2019MNRAS.487.2701O}. Instead of using some (poorly-constrained) forms for the 3- and 4-point correlation functions, we simply boost the level of small-scale power by rescaling the shot-noise via a factor $\alpha$, giving $\delta_i^2\rightarrow\frac{\alpha}{n_i}(1+\delta_i)$ and ignoring the $\xi^{(3)}_{ijk}$ and $\xi^{(4)}_{ijkl}$ terms. This corresponds to 2- and 3-cell summation terms in Eq.\,\ref{eq: n_point_expansion} by $\alpha^2$ and $\alpha$ respectively.

Inserting the definitions of Eq.\,\ref{eq: random_field_expansions} and the expansion Eq.\,\ref{eq: n_point_expansion} into $\av{C_{ab}^J}$, weighting the 2- and 3-point jackknife matrix expressions by $\alpha^2$ and $\alpha$, gives the simplified jackknife matrix expression of Eqs.\,\ref{eq: NewFullJackknifeC2-4Expression}\,\&\,\ref{eq: NewC234Definitions}. \resub{For faster integral sampling, we additionally replace $\xi_{ik}\xi_{jl}+\xi_{il}\xi_{jk}$ with $2\,\xi_{ik}\xi_{jl}$ in the jackknife matrix expressions, using the relabelling symmetry of indices in the integrals. The jackknife covariance terms (Eqs.\,\ref{eq: NewC234Definitions}) may be generalized into continuous space by replacing the summations by integrals and promoting all quantities to be functions of spatial position $\vec{r}$, giving}
\beq\label{eq: C234JackIntegrals}
    C_{4,ab}^J &=& \frac{1}{RR_a^JRR_b^J-\sum_BRR_{aB}RR_{bB}}\int{}d^3\vec{r}_id^3\vec{r}_jd^3\vec{r}_kd^3\vec{r}_l\,n(\vec{r}_i)n(\vec{r}_j)n(\vec{r}_k)n(\vec{r}_l)w(\vec{r}_i)w(\vec{r}_j)w(\vec{r}_k)w(\vec{r}_l)\Theta_a(\vec{r}_i-\vec{r}_j)\Theta_b(\vec{r}_k-\vec{r}_l)\\\nonumber
    &\times&\left[\xi^{(4)}(\vec{r}_i,\vec{r}_j,\vec{r}_k,\vec{r}_l)+\xi(\vec{r}_i-\vec{r}_j)\xi(\vec{r}_k-\vec{r}_l)+2\,\xi(\vec{r}_i-\vec{r}_k)\xi(\vec{r}_j-\vec{r}_l)\right]\omega^{ab}(\vec{r}_i,\vec{r}_j,\vec{r}_k,\vec{r}_l)\\\nonumber
    C_{3,ab}^J &=& \frac{4}{RR_a^JRR_b^J-\sum_BRR_{aB}RR_{bB}}\int{}d^3\vec{r}_id^3\vec{r}_jd^3\vec{r}_k\,n(\vec{r}_i)n(\vec{r}_j)n(\vec{r}_k)w(\vec{r}_i)w(\vec{r}_j)^2w(\vec{r}_k)\Theta_a(\vec{r}_i-\vec{r}_j)\Theta_b(\vec{r}_j-\vec{r}_k)\\\nonumber
    &\times&\left[\xi^{(3)}(\vec{r}_i,\vec{r}_j,\vec{r}_k)+\xi(\vec{r}_i-\vec{r}_k)\right]\omega^{ab}(\vec{r}_i,\vec{r}_j,\vec{r}_k,\vec{r}_j)\\\nonumber
    C_{2,ab}^J &=& \frac{2\delta_{ab}}{RR_a^JRR_b^J-\sum_BRR_{aB}RR_{bB}}\int{}d^3\vec{r}_id^3\vec{r}_j\,n(\vec{r}_i)n(\vec{r}_j)w^2(\vec{r}_i)w^2(\vec{r}_j)\Theta_a(\vec{r}_i-\vec{r}_j)\left[1+\xi(\vec{r}_i-\vec{r}_j)\right]\omega^{ab}(\vec{r}_i,\vec{r}_j,\vec{r}_i,\vec{r}_j)\\\nonumber
    RR_{aA} &=& \int{}d^3\vec{r}_id^3\vec r_j\,q^A(\vec r_i,\vec r_j)n(\vec{r}_i)n(\vec{r}_j)w(\vec{r}_i)w(\vec{r}_j)\Theta_a(\vec{r}_i-\vec{r}_j).
\eeq

\section{Disconnected Term Cancellation}\label{appen: disconnected}
As noted in Sec.\,\ref{subsec: disconnected_term}, the jackknife covariance matrix contains a disconnected term, which is a sum over two 2-point terms. Here, we consider the scenarios in which this unusual term cancels.

Consider a broad bin $a$, which can be split into a number of narrow sub-bins, $a'$, such that $\Theta_a^{ij} = \sum_{a'}\Theta_{a'}^{ij}$, where $\Theta^{ij}_{a'}$ are sufficiently narrow such that we can assume the 2PCF $\xi$ to take a constant value $\xi_{a'}$ across the bin. Noting that $q_{ij}^A$ is invariant of the bin $a'$ we have 
\beq
EE_{aA}-w_{aA}EE_a^J &=& \sum_{a'}\xi_{a'}\sum_{i\neq{}j}n_in_jw_iw_j\Theta^{ij}_{a'}\left(q_{ij}^{A}-w_{aA}Q_{ij}\right)\\\nonumber
&=&\sum_{a'}\xi_{a'}\left(RR_{a'A}-\frac{RR_{aA}}{\sum_CRR_{aC}}RR_{a'}^J\right)\\\nonumber
&=&\sum_{a'}\xi_{a'}\left(RR_{a'A}-\frac{\sum_BRR_{a'B}}{\sum_CRR_{aC}}RR_{aA}\right),
\eeq
inserting the full expression for the jackknife weights $w_{aA}$ and using the definition of $RR_{a'}^J$. We recover three conditions for cancellation of this term:
\begin{enumerate}
    \item $\xi_{a'}$ takes a constant value across sub-bins $a'$ in bin $a$. (This is the same as restricting the initial bin $a$ to be narrow such that $\xi_{ij}$ is constant across the bin). Cancellation occurs since $\sum_{a'}RR_{a'A} = RR_{aA}$ since the expressions are additive.
    \item $RR_{a'A}=RR_{a'B} \,\forall\,(A,B,a')$ i.e. all jackknife regions are identical (e.g. for a uniform survey). Here $\sum_{B}RR_{a'B} = N_JRR_{a'A}$ and $\sum_{B}RR_{aB}=N_JRR_{aA}$ which gives the desired cancellation.
    \item $RR_{a'A}$ is independent of the sub-bin $a'$, i.e. we have the same pair counts (and not necessarily uniform $\xi_{ij}$) across each of the $N_\mathrm{sub}$ sub-bins in the full bin $a$. Here $RR_{a'A} = \frac{1}{N_\mathrm{sub}}RR_{aA}$ gives the desired cancellation.
\end{enumerate}
If these simplifying conditions are not met, as would generally be the case for a realistic survey geometry and 2PCF, the disconnected term is non-cancelling, although we expect it to be small.

\section{Probability Grid Integrals}\label{appen: probability_integrals}
We here derive the approximate solutions to the probability sampling grid integral (Eq.\,\ref{eq: probIntegralInitial}); 
\beq\label{eq: probIntegralInitial_appen}
    A(\vec{n})=\intop\intop W_{\vec{0}}\left(\vec{y}\right)W_{\vec{n}}\left(\vec{x}\right)K\left(\vec{x}-\vec{y}\right)d^{3}\vec{x}d^{3}\vec{y}\;.
\eeq
First we note that this may be rewritten as a convolution (denoted by a $\star$):
\beq
A(\vec{n}) &=& \int W_\vec{n}(\vec{x})[W_\vec{0}\star{}K](\vec{x})d^3\vec{x}\\\nonumber
&=& \left[\left(W_\vec{0}\star{}K\right)\star{}W_\vec{0}\right](\vec{n})
\eeq
where we have used $W_\vec{n}(\vec{x}) = W_\vec{0}(\vec n-\vec{x})$. Applying the convolution theorem in Fourier space gives 
\beq
\tilde{A}(\vec{k}) &=& \tilde{W}_\vec{0}^2(\vec{k})\tilde{K}(\vec{k})\\\nonumber
\Rightarrow A(\vec{n}) &=& \left[\mathcal{F}^{-1}\left(\tilde{W}_\vec{0}^2\right)\star{}K\right](\vec{n}).
\eeq
The first simplifying assumption is to assume spherical rather than cubic symmetry of the window functions (keeping the volume $V = a^3$ fixed) such that
\beq W_{\vec{0},\mathrm{TH}}(\vec{x}) = \begin{cases}\frac{1}{V}&\text{ if }|\vec{x}|<r_0 = \left(\frac{3V}{4\pi}\right)^{1/3}\\ 0 &\text{ else.}\end{cases}
\eeq
This has the standard Fourier transform
\beq
\tilde{W}_{\vec{0},\mathrm{TH}}(\vec{k}) = \frac{3}{r_0k}j_1(r_0k)
\eeq
for $k = |\vec{k}|$ and first-order spherical Bessel function $j_1$. Second, we can approximate this as a Gaussian window function of width $R = a/2$, which will give a more tractable integral. This has functional form 
\beq
W_{\vec{0},\mathrm{G}}(\vec{x}) = \frac{1}{(\sqrt{2\pi}R)^3}e^{-\frac{r^2}{2R^2}}
\eeq
with Fourier transform
\beq
\tilde{W}_{\vec{0},\mathrm{G}}(\vec{k}) = e^{-\frac{k^2R^2}{2}}.
\eeq
For a spherical-top hat window function, we can compute the inverse Fourier transform of $\tilde{W}_{\vec{0},\mathrm G}^2$:
\beq
g_\mathrm{TH}(n) \equiv \mathcal{F}^{-1}\left[\tilde{W}_{\vec{0},\mathrm{TH}}^2\right](\vec{n}) &=& \frac{1}{(2\pi)^3}\frac{9}{r_0^2}\int{}e^{-i\vec{k}\cdot\vec{n}}\frac{1}{k^2}j_1^2(r_0k)d^3\vec{k}\\\nonumber
&=& \frac{3}{(2\pi{}r_0)^2}\int_0^\infty{}dk\int_{-1}^{1}d\mu e^{-ikn\mu}j_1^2(r_0k)\\\nonumber
&=& \frac{1}{2}\left(\frac{3}{\pi{}r_0}\right)^2\int_0^\infty{}dkj_0(nk)j_1^2(r_0k).
\eeq
For a spherical Gaussian window:
\beq
g_\mathrm{G}(n) \equiv \mathcal{F}^{-1}\left[\tilde{W}_{\vec{0},\mathrm{G}}^2\right](\vec{n})&=& \frac{1}{(2\pi)^3}\int d^3\vec{k} e^{-i\vec{k}\cdot\vec{n}}e^{-k^2R^2}\\\nonumber
&=& \frac{1}{2\pi^2n}\int_0^\infty \sin(kn)e^{-k^2R^2}k\,dk\\\nonumber
&=& \frac{1}{8\pi^{3/2}R^3}\exp\left(-\frac{n^2}{4R^2}\right)
\eeq
\citep{gradshteyn2007}. This allows computation of $A(\vec{n}) = \left[g\star{}K\right](\vec{n})$, which reduces to a 2D integral by symmetry:
\beq
A(n) = 2\pi\int_0^\infty m^2dm \int_{-1}^1d\mu K(m)g\left(\left[m^2+n^2-2mn\mu\right]^{1/2}\right).
\eeq
With the Gaussian window function further simplifications are possible;
\beq\label{eq: probIntegralForXiAppen}
A_\mathrm{G}(n) &=& 2\pi\int_0^\infty m^2\,dm\int_{-1}^1d\mu \frac{K(m)}{8\pi^{3/2}R^3}\exp\left(-\frac{m^2+n^2-2mn\mu}{4R^2}\right)\\\nonumber
&=&\frac{1}{Rn\sqrt{\pi}}\exp\left(-\frac{n^2}{4R^2}\right)\int_0^\infty m\,dm K(m) \exp\left(-\frac{m^2}{4R^2}\right)\sinh\left(\frac{mn}{2R^2}\right).
\eeq
This may be computed numerically for arbitrary $K(m)$. In the limit $n\rightarrow0$ (corresponding to selecting the same cell), we use a Taylor series expansion to define
\beq
A_\mathrm{G}(0) = \frac{1}{2\sqrt{\pi}R^3}\int_0^\infty m^2\,dm K(m)\exp\left(-\frac{m^2}{4R^2}\right).
\eeq
These are the forms used in this paper to compute $A(n)$ for the $K(\vec r) = \xi(|\vec r|)$ kernel. An alternative approach involves the direct inverse Fourier transform on the product $\tilde{W}_\vec{0}^2\tilde{K}$, which depends only on $k = |\vec{k}|$. For a top-hat window function:
\beq
A_\mathrm{TH}(n) &=&\frac{1}{(2\pi)^3}\int d^3\vec{k}e^{-i\vec{k}\cdot\vec{n}}\left(\frac{3}{r_0k}\right)^2j_1^2(r_0k)\tilde{K}(k)\\\nonumber
&=& \frac{3}{(2\pi{}r_0)^2}\int_0^\infty{}dk\int_{-1}^{1}d\mu e^{-ikn\mu}j_1^2(r_0k)\tilde{K}(k)\\\nonumber
&=& \frac{1}{2}\left(\frac{3}{\pi r_0}\right)^2\int_0^\infty dk j_0(nk) j_1^2(r_0k)\tilde{K}(k)
\eeq
which can be computed numerically, if $\tilde{K}(k)$ is known (e.g. $\tilde{K}(k) = P(k)$ (the one-dimensional galaxy-galaxy power spectrum) for $K(\vec r) = \xi(|\vec r|)$). For a spherical Gaussian window function we have the alternate form:
\beq
A_\mathrm{G}(n) = \frac{1}{2\pi^2}\int_0^\infty j_0(kn)e^{-k^2R^2}\tilde{K}(k)k^2dk.
\eeq
Specializing to the $1/r^2$ kernel, we have Fourier transform
\beq
\tilde{K}(k) = \int d^3\vec{r}\frac{e^{i\vec{k}\cdot\vec{r}}}{r^2} = 4\pi\int_0^\infty j_0(kr) dr = \frac{2\pi^2}{k}. 
\eeq
Using the Gaussian window function permits the semi-analytic result
\beq\label{eq: probIntegralForR2Appen}
A_\mathrm{G}(n) = \frac{2}{an}F\left(\frac{n}{a}\right)
\eeq
\citep{gradshteyn2007}, where $F$ is the Dawson-F function \citep{DawsonF}. As $r\rightarrow0$, we recover $A_\mathrm G(n) \rightarrow2/{a^2}$ via a Taylor expansion.

\section{Relating the KL Divergence to Sample Size}\label{appen: KL-div}
We proceed to derive a useful result relating the KL divergence between two noisy matrices $\{X_i\}$ to their number of matrix samples, $\{n^{(i)}_s\}$. For a general noisy sample covariance matrix $X$ computed from $n_s$ draws of a multivariate normal distribution with covariance (precision) matrix $C_0$ ($\Psi_0$), the noise on $X$ is Wishart distributed, with expected covariance
\beq\label{eq: wishart_cov}
    \operatorname{cov}(X_{ab},X_{cd}) = \frac{C_{0,ac}C_{0,bd}+C_{0,ad}C_{0,bc}}{n_s}.
\eeq
Decomposing $X_i = C_0+\delta X_i$ we can expand
\beq
    X_1^{-1} &=& \Psi_0 \left(1+\Psi_0\delta X_1\right)^{-1}\\\nonumber
    &=& \Psi_0 \left(1 -\Psi_0\delta X_1 + \Psi_0\delta X_1 \Psi_0 \delta X_1\right)+\mathcal{O}(\delta X_1^3) 
\eeq
thus 
\beq
\mathrm{tr}(X_1^{-1},X_2) =  \mathrm{tr}\left[\Psi_0\left(1-\Psi_0\delta X_1 + \Psi_0\delta X_1 \Psi_0 \delta X_1\right)\left(C_0+\delta X_2\right)\right].
\eeq
Taking the expectation, and assuming $\delta X_1$ and $\delta X_2$ are independent, such that $\av{\delta X_1\delta X_2} = 0$, we obtain
\beq
    \av{\mathrm{tr}(X_1^{-1},X_2)}  = n_\mathrm{bins}+\mathrm{tr}(\Psi_0\av{\delta X_1\Psi_0\delta X_1}) 
\eeq
using $\av{\delta X_1}=\av{\delta X_2} = 0$ and $\Psi_0C_0 = \mathbb{I}$. Inserting this into the KL divergence expectation (Eq.\,\ref{eq: KL_divergence}) gives
\beq
    2\av{D_{KL}(X_1^{-1},X_2)} = \mathrm{tr}(\Psi_0\av{\delta X_1 \Psi_0 \delta X_1})-\av{\log\det X_1^{-1}} - \av{\log\det X_2}.
\eeq
To simplify this further, first note that $\log\det X_1^{-1} = -\log\det X_1$ and
\beq\label{eq: KL_tmp}
    \log\det X_i &=& \log\left[\det(C_0)\det(1+\Psi_0\delta X_i)\right]\\\nonumber
    &=& \log\det C_0 + \log\det(1+\Psi_0\delta X_i).
\eeq
Using the identity $\log\det Y \equiv \mathrm{tr}\log{Y}$, we can expand Eq.\,\ref{eq: KL_tmp} to quadratic order in $\delta X_i$;
\beq
    \av{\log\det X_i} \approx \log\det C_0 - \frac{1}{2}\mathrm{tr}(\Psi_0\av{\delta X_i \Psi_0 \delta X_i}),
\eeq
which, when inserted into the KL divergence expectation, gives the form
\beq
    2\av{D_{KL}(X_1^{-1},X_2)} \approx \frac{1}{2}\left[\mathrm{tr}(\Psi_0\av{\delta X_1\Psi_0\delta X_1})+\mathrm{tr}(\Psi_0\av{\delta X_2\Psi_0\delta X_2)}\right].
\eeq
Via the Wishart covariance expansion (Eq.\,\ref{eq: wishart_cov}), we can expand the traces as
\beq
    \mathrm{tr}(\Psi_0\av{\delta X_i \Psi_0\delta X_i}) &=& \Psi_{0,ab}\av{\delta X_{i,bc}\Psi_{0,cd}\delta X_{i,da}}\\\nonumber
    &=& \frac{1}{n_s^{(i)}}\left(\Psi_{0,ab}C_{0,bd}\Psi_{0,cd}C_{0,da}+\Psi_{0,ab}C_{0,ba}\Psi_{0,cd}C_{0,dc}\right) = \frac{n_\mathrm{bins}(n_\mathrm{bins}+1)}{n_s^{(i)}},
\eeq
giving the general result
\beq\label{eq: relating_KL_div_to_n_samples_general}
    \av{D_{KL}(X_1^{-1},X_2)}
    \approx\frac{n_\mathrm{bins}(n_\mathrm{bins}+1)}{4}\left(\frac{1}{n_s^{(1)}}+\frac{1}{n_s^{(2)}}\right)
\eeq
valid in the limit $n_s\gg n_\mathrm{bins}$. This is of particular importance when one matrix is far smoother than the other, such that $n^{(1)}_s\gg n^{(2)}_s=n_s$ and $X_1^{-1}\approx \Psi_0$. In this case
\beq\label{eq: relating_KL_div_to_n_samples_simple}
    \av{D_{KL}(\Psi_0,X_2)}
    \approx\frac{n_\mathrm{bins}(n_\mathrm{bins}+1)}{4n_s}
\eeq
This gives the desired conversion between the KL divergence of a noisy draw from a multivariate distribution with some known smooth precision matrix $\Psi_0$ and the number of samples, $n_s$, which is approximated by $n_\mathrm{eff}(X_2)$ for non-Wishart noise.


\bsp	
\label{lastpage}
\end{document}